\documentclass[12pt,preprint]{aastex}

\usepackage{ulem}
\usepackage{longtable}
\usepackage{lscape}
\def\meth {CH$_3$OH}
\def\hho  {H$_2$O}

\def\kms  {km~s$^{-1}$}

\def\masy {mas~y$^{-1}$}

\def\deg  {\ifmmode {^\circ}\else {$^\circ$}\fi}
\def\porm {\ifmmode {\pm}\else {$\pm$}\fi}
\def\chisqpdf {\ifmmode {\chi^2_{\rm pdf}}\else {$\chi^2_{\rm pdf}$}\fi}
\def\chisq    {\ifmmode {\chi^2}\else {$\chi^2$}\fi}

\def\d    {\ifmmode {{\rlap{.}}^\circ}\else {${\rlap{.}}^\circ$}\fi}
\def\s    {\ifmmode {{\rlap{.}}^s}\else {${\rlap{.}}^s$}\fi}
\def\as   {\ifmmode {{\rlap{.}}^{''}}\else {${\rlap{.}}^{''}$}\fi}

\def\pa    {\ifmmode {\psi} \else {$\psi$}\fi}

\def\vlsr  {\ifmmode {v_{\rm LSR}}\else {$v_{\rm LSR}$}\fi}
\def\vlsrr {\ifmmode {v^r_{\rm LSR}}\else {$v^r_{\rm LSR}$}\fi}
\def\vhelio{\ifmmode {v_{Helio}}\else {$v_{Helio}$}\fi}

\def\ura   {\ifmmode {\mu_\alpha}\else {$\mu_\alpha$}\fi}
\def\udec  {\ifmmode {\mu_\delta}\else {$\mu_\delta$}\fi}

\def\ul    {\ifmmode {\mu_l}\else {$\mu_l$}\fi}
\def\ub    {\ifmmode {\mu_b}\else {$\mu_b$}\fi}

\def\uml   {\ifmmode {v_{gr}}\else {$v_{gr}$}\fi}
\def\umb   {\ifmmode {v_b}\else {$v_b$}\fi}
\def\vsrad {\ifmmode {v_{rad}}\else {$v_{rad}$}\fi}

\def\upl   {\ifmmode {v^p_{gr}}\else {$v^p_{gr}$}\fi}
\def\upb   {\ifmmode {v^p_b}\else {$v^p_b$}\fi}
\def\vprad {\ifmmode {v^p_{rad}}\else {$v^p_{rad}$}\fi}

\def\Vo    {\ifmmode {V^{Std}_\odot}\else {$V^{Std}_\odot$}\fi}
\def\Uo    {\ifmmode {U^{Std}_\odot}\else {$U^{Std}_\odot$}\fi}
\def\Wo    {\ifmmode {W^{Std}_\odot}\else {$W^{Std}_\odot$}\fi}
\def\VH    {\ifmmode {V^H_\odot}\else {$V^H_\odot$}\fi}
\def\UH    {\ifmmode {U^H_\odot}\else {$U^H_\odot$}\fi}
\def\WH    {\ifmmode {W^H_\odot}\else {$W^H_\odot$}\fi}
\def\V     {\ifmmode {V_\odot}\else {$V_\odot$}\fi}
\def\U     {\ifmmode {U_\odot}\else {$U_\odot$}\fi}
\def\W     {\ifmmode {W_\odot}\else {$W_\odot$}\fi}

\def\Vs    {\ifmmode {V_s}\else {$V_s$}\fi}
\def\Us    {\ifmmode {U_s}\else {$U_s$}\fi}
\def\Ws    {\ifmmode {W_s}\else {$W_s$}\fi}

\def\Vsbar {\ifmmode {\overline{V_s}}\else {$\overline{V_s}$}\fi}
\def\Usbar {\ifmmode {\overline{U_s}}\else {$\overline{U_s}$}\fi}
\def\Wsbar {\ifmmode {\overline{W_s}}\else {$\overline{W_s}$}\fi}

\def\pars  {\ifmmode{\pi_s}\else{$\pi_s$}\fi}

\def\Ts    {\ifmmode{\Theta_s}\else{$\Theta_s$}\fi}
\def\Tdot  {\ifmmode{d\Theta\over dR}\else{$d\Theta\over dR$}\fi}

\def\Rp    {\ifmmode{R_p}\else{$R_p$}\fi}

\def\To    {\ifmmode{\Theta_0}\else{$\Theta_0$}\fi}
\def\Ro    {\ifmmode{R_0}\else{$R_0$}\fi}

\def\Gsrc {G059.78$+$00.06}
\def\Ge {G074.03$-$01.71}
\def\Gf {G075.76$+$00.33}
\def\Gm {G075.78$+$00.34}
\def\Gg {G076.38$-$00.61}
\def\Gk {G079.87$+$01.17}
\def\Gh {G090.21$+$02.32}
\def\Gi {G092.67$+$03.07}

\def\Gl {G105.41$+$09.87}

\def\Vlsr {\ifmmode {V_{\rm LSR}} \else {$V_{\rm LSR}$} \fi}

\slugcomment{Draft: 2013 March 28}

\shorttitle{Distance to the Local arm}
\shortauthors{Xu et al.}

\begin{document}

\title{ON THE NATURE OF THE LOCAL SPIRAL ARM OF THE MILKY WAY}

\author{Y. Xu\altaffilmark{1}, J. J. Li\altaffilmark{1,2}, M. J.
Reid\altaffilmark{3}, K. M.
Menten\altaffilmark{2}, X. W. Zheng\altaffilmark{4}, A. Brunthaler\altaffilmark{2},
L. Moscadelli\altaffilmark{5}, T. M. Dame\altaffilmark{3} and B.
Zhang$^2$}

\altaffiltext{1}{Purple Mountain Observatory, Chinese Academy of
Sciences, Nanjing 210008, China; xuye@pmo.ac.cn}
\altaffiltext{2}{Max-Planck-Institut f$\ddot{u}$r
Radioastronomie, Auf dem H{\" u}gel 69, 53121 Bonn, Germany}
\altaffiltext{3}{Harvard-Smithsonian Center for Astrophysics,
60 Garden Street, Cambridge, MA 02138, USA}
\altaffiltext{4}{Nanjing University, Nanjing 20093, China}
\altaffiltext{5}{INAF-Osservatorio Astrofisico di Arcetri, Largo E.
Fermi 5, 50125 Firenze, Italy}

\begin{abstract}

Trigonometric parallax measurements of nine water masers associated
with the Local arm of the Milky Way were carried out as part of the BeSSeL Survey using
the VLBA. When combined with 21 other parallax measurements from the literature,
the data allow us to study the distribution and 3-dimensional motions of star
forming regions in the spiral arm over the entire northern sky.
Our results suggest that the Local arm does not have the large pitch angle
characteristic of a short spur.  Instead its active star formation, overall length ($>5$ kpc),
and shallow pitch angle ($\sim$10$^{\circ}$) suggest that it is more
like the adjacent Perseus and Sagittarius arms; perhaps it is a branch of one of these arms.
Contrary to previous results, we find the Local arm to be closer to the Perseus than
to the Sagittarius arm, suggesting that a branching from the former may be more likely.
An average peculiar motion of near-zero toward both the Galactic center and north
Galactic pole, and counter rotation of $\sim$ 5 \kms\ were observed, indicating
that the Local arm has similar kinematic properties as found for other major spiral arms.

\end{abstract}

\section{INTRODUCTION}

Determining the spiral structure of our Milky Way has been
a long-standing problem in astrophysics, since pioneering studies by
\citet{Mor:52,Mor:53} revealed three spiral arm segments in the solar
neighborhood by spectroscopic parallax. However, progress has been hampered by the
difficulty in identifying spectroscopic distances for distant
sources, because of large and variable extinction from interstellar dust in the
galactic plane. Although distances derived with other methods, such
as kinematic distances, offer opportunities for progress, as
demonstrated by \citet{Georgelin:76} (GG76 hereafter) and
\citet{Rus:07}, typical uncertainties are comparable to the spacing
between arms and preclude identification of spiral arms of the Milky
Way with confidence. At present, there is no general agreement
on the number of arms nor on their locations and orientations. \citet{Churchwell:09}
favor two arms from their investigation of counts of older stars,
while observations of gas and dust indicate four- or five-arm
models \citep{Rus:03,Hou:09}. Many dozens of models have
been proposed (see the review of \citet{Liszt:85,Steiman:10}),
most of which are four-armed structures similar to that of GG76.

The major spiral arms within a few kpc of the Sun (Sagittarius
and the Perseus arms) are well known, appearing as arcs in longitude-velocity
plots of HI and CO emission \citep{Oort:58,Bok:59,Georgelin:76,Dame:01}.
Parallaxes for high-mass star-forming regions (HMSFRs)
now clearly locate the Sagittarius and the Perseus arms in the Galaxy
with 3-dimensional positions and velocities (Choi, in preparation; Wu, in preparation).
However, star-forming regions between these arms are known and collectively
they have been called the ``Orion spur'', the ``Orion arm'', or the
``Local arm'' \citep{Bok:70,Kerr:70,Georgelin:76,Avedisova:85}. Generally the
Local arm was thought to be a secondary spiral feature, because the
density of star-forming regions was significantly less than
that of other major arms.

As part of the Bar and Spiral Structure Legacy (BeSSeL) Survey, we have
been measuring trigonometric parallaxes for large numbers of HMSFRs. Somewhat surprisingly, we are also finding many
sources thought to be in the Perseus arm (from kinematic distances) are
instead associated with the Local arm. For example, we found that the
HMSFR \Gf, which has a kinematic distance of $\approx5.7$~kpc and would locate it in the Perseus arm, has a
distance measured by trigonometric parallax of $3.5\pm0.3$~kpc,
which places it in the Local arm (Section~\ref{result}).
Based on BeSSeL Survey, VLBI Exploration of Radio Astrometry (VERA),
and European VLBI Network (EVN) parallaxes, we can now identify
30 sources that are associated with the Local arm. Using these accurate
distances and proper motions, we can now study the geometry and kinematics
of the Local arm with unprecidented detail.

\section{OBSERVATIONS}

Observations of 22-GHz \hho\ masers towards star-forming regions
were carried out with the National Radio Astronomy Observatory
(NRAO)\footnote{The National Radio Astronomy Observatory is a
facility of the US National Science Foundation operated under
cooperative agreement with Associated Universities, Inc.} Very Long
Baseline Array (VLBA), under programs BR145 and BM272.
All sources were observed over 7- or 9-h tracks
for projects BR145 or BM272, respectively.
Table~\ref{table:observations} lists details for the epochs
observed. The observations consisted of three (BM272) or four (BR145)
geodetic blocks, with phase-referenced observations inserted between
these blocks. Details of the observational setup and calibration
procedures can be found in \citet{Reid:09a}. All data were processed
on the DiFX\footnote{This work made use of the Swinburne University of Technology
software correlator, developed as part of the Australian Major National
Research Facilities Programme and operated under licence.} correlator in Socorro, NM \citep{Del:07}.

\begin{deluxetable}{llccccccc}
\tabletypesize{\tiny} \tablecaption{Details of the Epochs Observed}
\tablehead{\colhead{Code} & \colhead{Source} & \colhead{Epoch~1}
& \colhead{Epoch~2} & \colhead{Epoch~3} & \colhead{Epoch~4} & \colhead{Epoch~5}
& \colhead{Epoch~6} & \colhead{Epoch~7}}
\startdata
 BM272 & ON 1 & 2008 Nov 11 & 2009 May 7 & 2009 Nov 14 & 2009 Dec
20 & 2010 May 5 & 2011 Feb 26 & \nodata \\
 & \Gm & \nodata & 2009 May 7 & \nodata & 2009 Dec 20 & 2010 May
5 & 2011 Feb 26 & \nodata \\
 BR145D & \Ge & 2010 Apr 24 & 2010 Jul 26 & 2010 Sep 25 & 2010 Nov
18 & 2010 Dec 16 & 2011 Apr 25 & \nodata \\
 & \Gf & 2010 Apr 24 & 2010 Jul 26 & 2010 Sep 25 & 2010 Nov 18 &
2010 Dec 16 & 2011 Apr 25 & \nodata \\
 & \Gg & 2010 Apr 24 & 2010 Jul 26 & 2010 Sep 25 & 2010 Nov 18 &
2010 Dec 16 & 2011 Apr 25 & \nodata \\
 BR145E & \Gh & 2010 May 16 & 2010 Aug 9 & 2010 Oct 12 & 2010 Nov
27 & 2010 Dec 31 & 2011 May 13 & \nodata \\
 & \Gi & 2010 May 16 & 2010 Aug 9 & 2010 Oct 12 & 2010 Nov 27 & 2010
Dec 31 & 2011 May 13 & \nodata \\
 BR145S & \Gk & 2011 May 24 & 2011 Aug 8 & 2011 Oct 30 & 2011 Nov
26 & 2012 Jan 12 & 2012 May 14 & 2012 Jun 29\\
 & \Gl & 2011 May 24 & 2011 Aug 8 & 2011 Oct 30 & 2011 Nov 26 & 2012
Jan 12 & 2012 May 14 & 2012 Jun 29\\
\enddata
\label{table:observations}
\end{deluxetable}

For each maser source, we used several different background sources,
selected from the VCS2 and VCS3 catalogs \citep{Fomalont:03,
Petrov:05} and our VLBI calibrator surveys \citep{Xu:06a, Immer:11}.
\hho\ masers can be time-variable with lifetimes of months to years.
By comparison background compact extragalactic radio sources are
relatively stable, so we used them as phase references when
they were strong enough to do so. After making maps of both maser and background sources for all
epochs, we measured the source positions and brightnesses by fitting
elliptical Gaussian brightness distributions.
Table~\ref{table:positions} presents position and intensity data for
the masers used for parallax measurements and the corresponding
background sources, as well as other observational parameters.
Absolute maser positions are derived from the positions of the
corresponding VCS or ICRF sources \citep{Ma:98}, which are generally
accurate within $\approx1$~mas.

\section{RESULTS}\label{result}

Here we focus on parallax and absolute proper motion measurements; images of
maser spots, background extragalactic continuum sources and internal maser
motions are presented in the Online Material. Parallax and
proper motion were fitted simultaneously to the data. Because
systematic errors (owing to small uncompensated atmospheric delays
and, in some cases, varying maser and calibrator source structures)
typically dominate over thermal noise when measuring relative source
positions, we added ``error floors'' in quadrature to the formal
position uncertainties. We used different error floors for the Right
Ascension and Declination data and adjusted them to yield post-fit
residuals with $\chi^2$ per degree of freedom near unity for both
coordinates.

\begin{figure}
\includegraphics[angle=-90,scale=0.6]{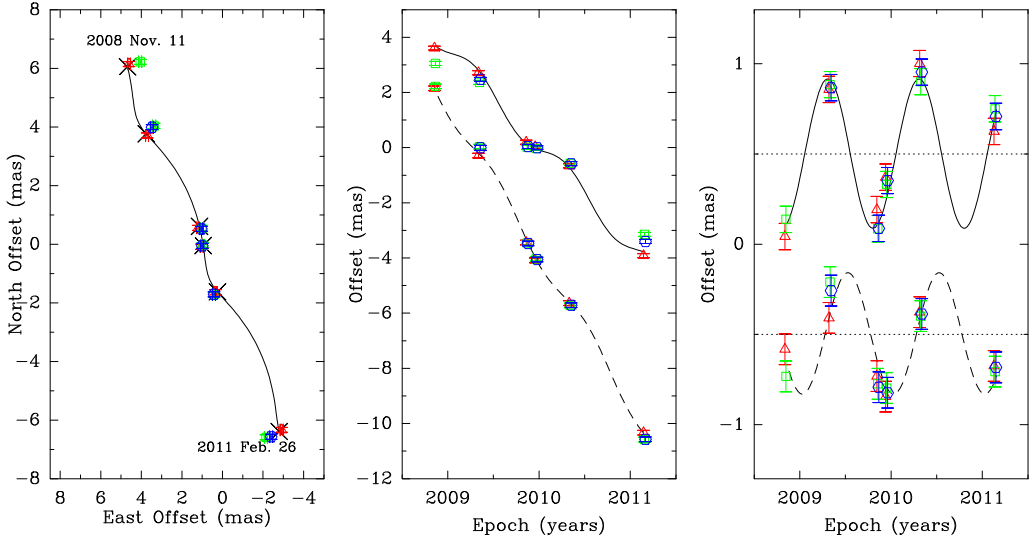}
\caption{\scriptsize Parallax and proper motion data and fits for ON~1. Plotted
are position offsets for one maser spot at \Vlsr = 14.8 (red
triangles) and two maser spots at 10.6~\kms\ (green squares for the
strong spot and blue hexagons for the weak spot) relative to the
background source J2003$+$3034. {\it Left Panel:} Positions in the
sky with the first and last epochs labeled. The expected positions
from the parallax and proper motion fit are indicated (crosses).
{\it Middle Panel:} East (solid line) and north (dashed line)
position offsets and best parallax and proper motions fits versus
time. {\it Right Panel:} Same as the {\it middle panel} but with the
best-fit proper motions removed for clearer illustration of the
parallax sinusoid.\label{on1_parallax}}
\end{figure}

\begin{figure}
\includegraphics[angle=-90,scale=0.6]{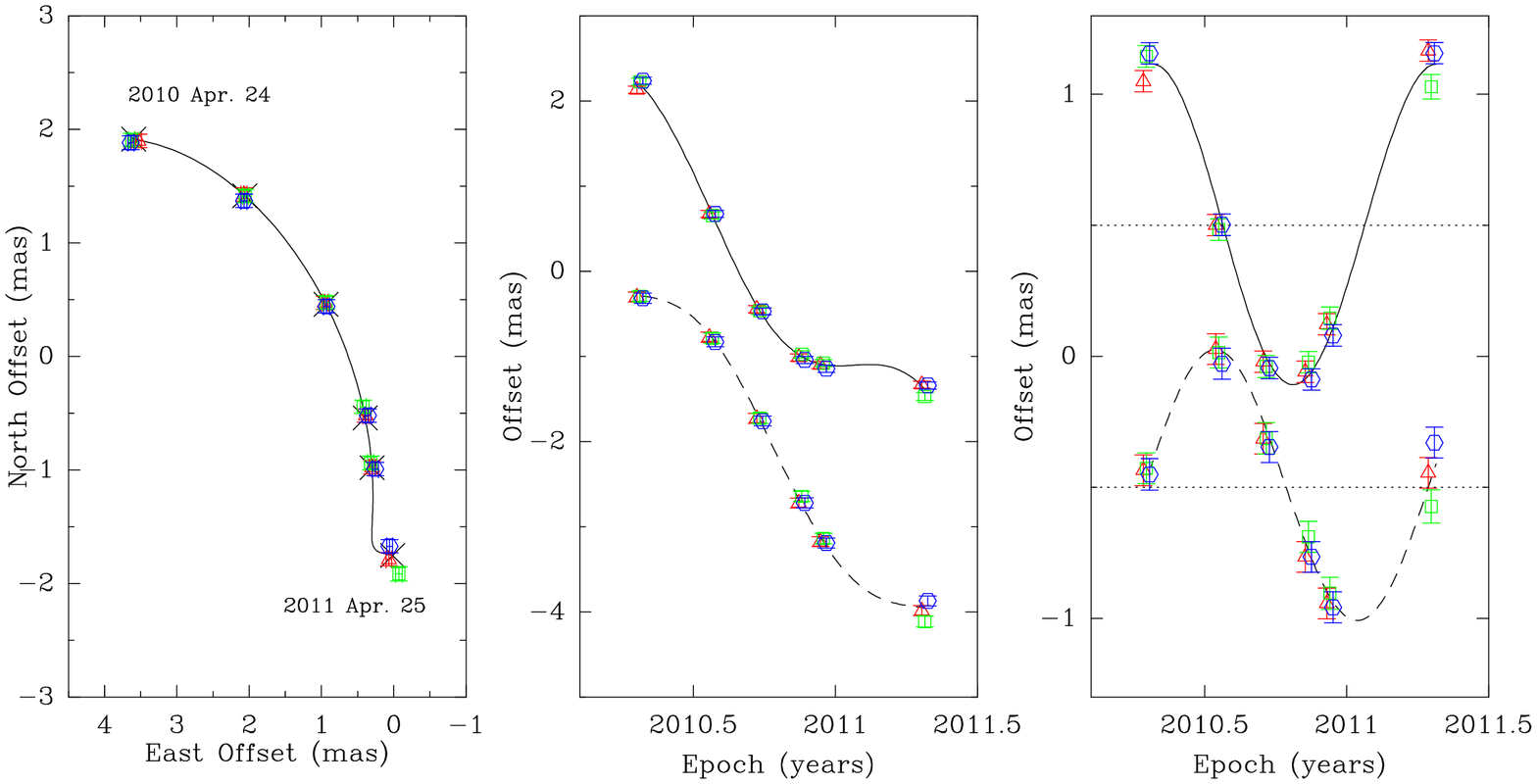}
\caption{\scriptsize Parallax and proper motion data and fits for \Ge.\ Plotted
are position offsets for three maser spots at \Vlsr = 13.4 (red
triangles), 12.9 (green square) and 12.5~\kms\ (blue hexagons)
relative to the background source J2025$+$3343. Solid and dashed
lines in the three panels represent the same fits as in the
corresponding panels of Fig.~\ref{on1_parallax}.
\label{g74_parallax}}
\end{figure}

\begin{figure}
\includegraphics[angle=-90,scale=0.6]{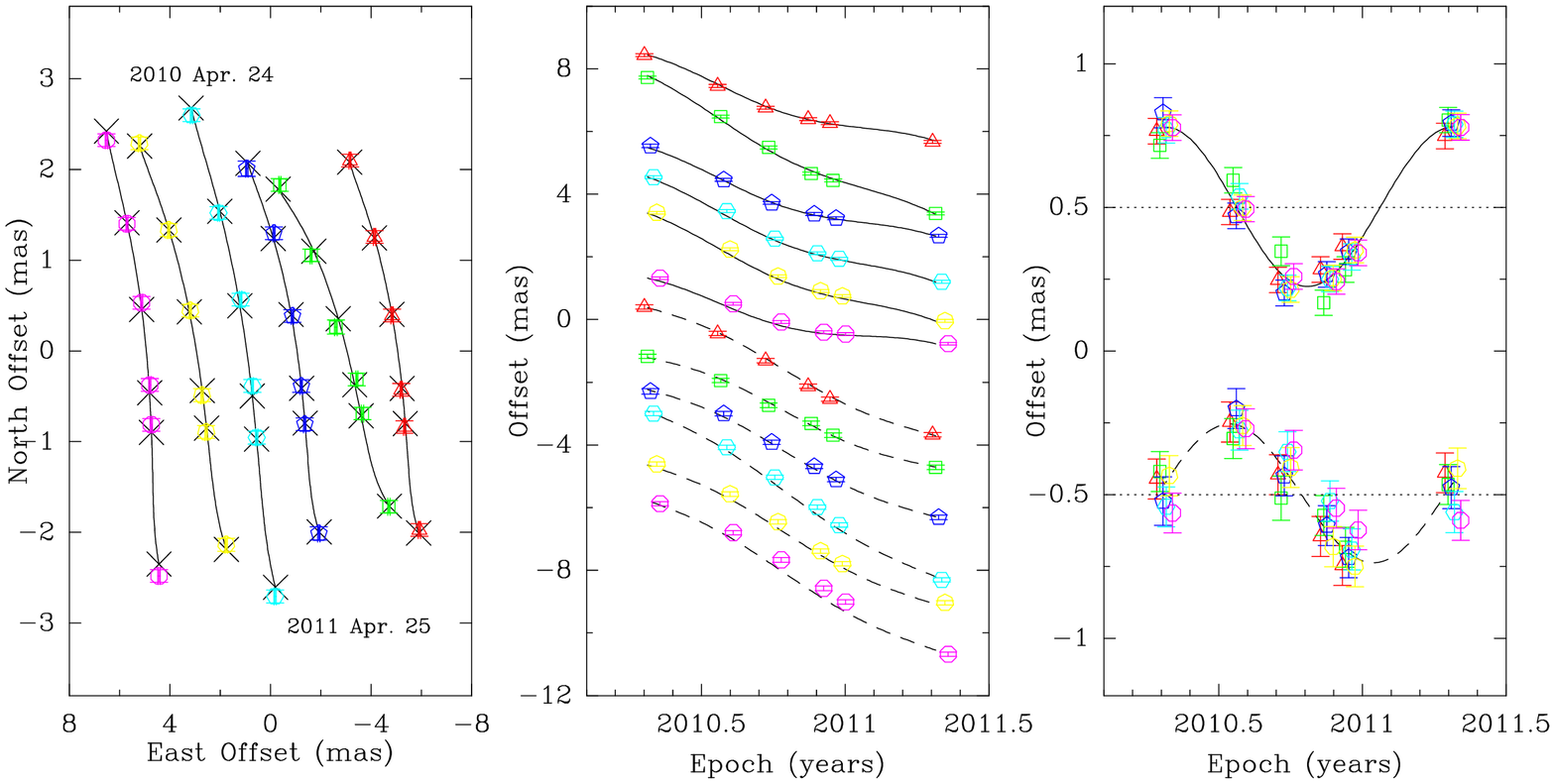}
\caption{\scriptsize Parallax and proper motion data and fits for \Gf. Plotted
are position offsets of 6 maser spots at \Vlsr = -5.0 (red
triangles), -7.1 (green square), -9.2 (blue pentagons), -9.6 (cyan
hexagons), -10.4 (yellow heptagons), and -11.3~\kms\ (fuchsia
octagons) relative to the background source J2015$+$3710. Solid and
dashed lines in the three panels represent the same fits as in the
corresponding panels of Fig.~\ref{on1_parallax}.
\label{g75_parallax}}
\end{figure}

\begin{figure}
\includegraphics[angle=-90,scale=0.6]{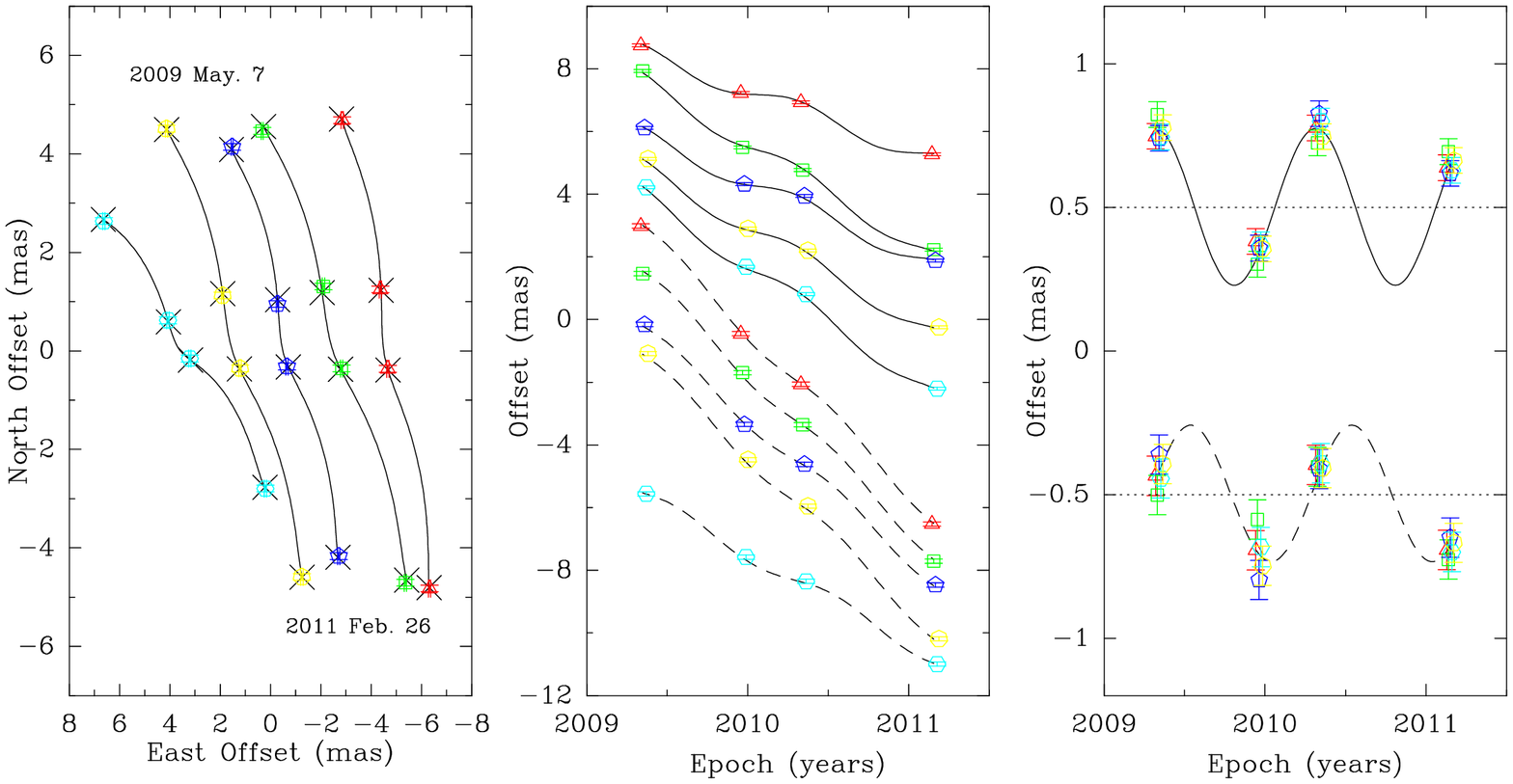}
\caption{\scriptsize Parallax and proper motion data and fits for \Gm. Plotted
are position offsets of 5 maser spots at \Vlsr = 0.4 (red
triangles), 2.9 (green square), 3.4 (blue pentagons), 3.8 (cyan
hexagons), and 4.2 (yellow heptagons) relative to the background
source J2015$+$3710. Solid and dashed lines in the three panels
represent the same fits as in the corresponding panels of
Fig.~\ref{on1_parallax}. \label{g75bm272_parallax}}
\end{figure}

\begin{figure}
\includegraphics[angle=-90,scale=0.6]{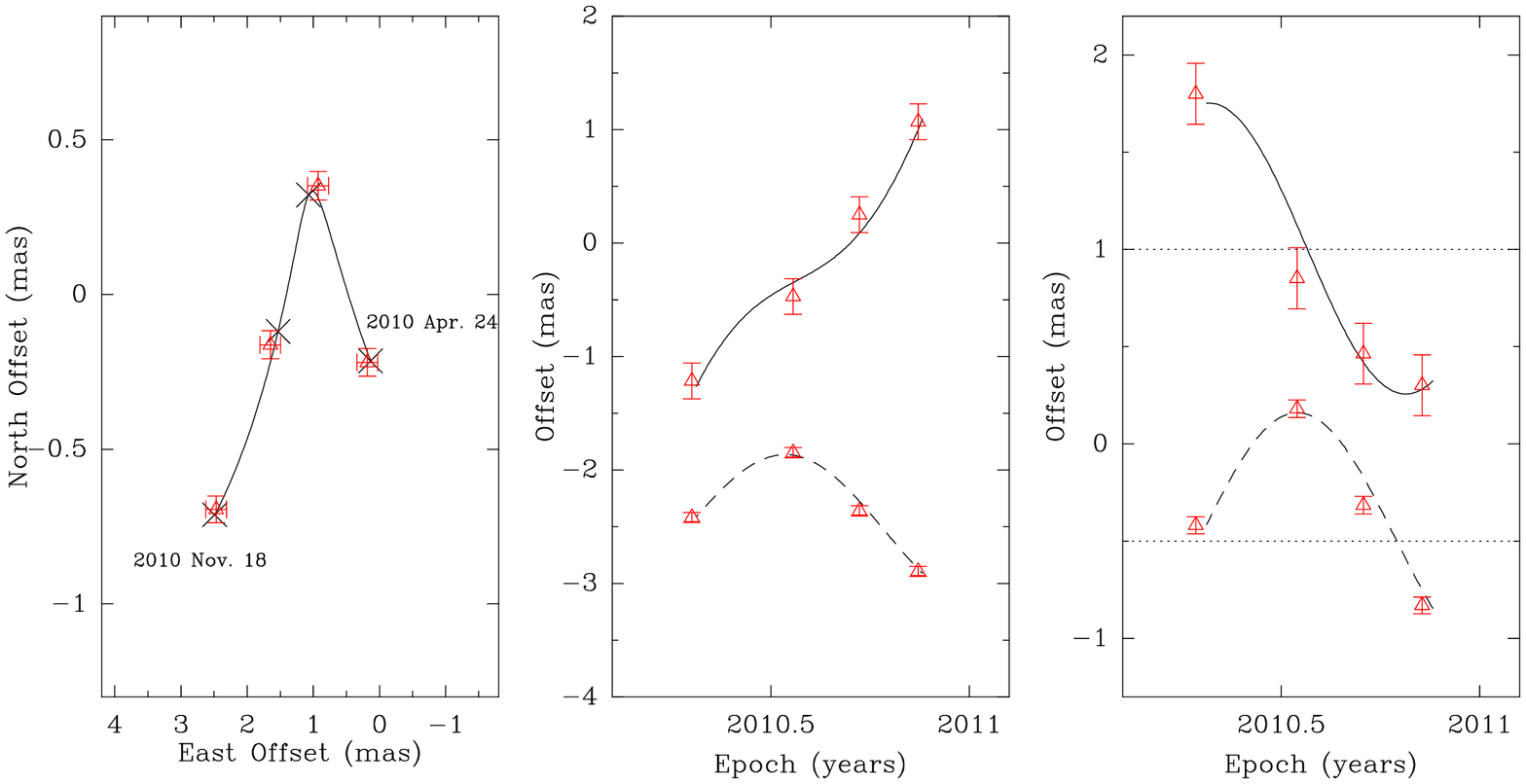}
\caption{\scriptsize Parallax and proper motion data and fits for \Gg. Plotted
are position offsets of the maser spot at \Vlsr = 6.9~\kms\ (red
triangles) relative to the background source J2015$+$3710. Solid and
dashed lines in the three panels represent the same fits as in the
corresponding panels of Fig.~\ref{on1_parallax}.
\label{g76_parallax}}
\end{figure}

\begin{figure}
\includegraphics[angle=-90,scale=0.6]{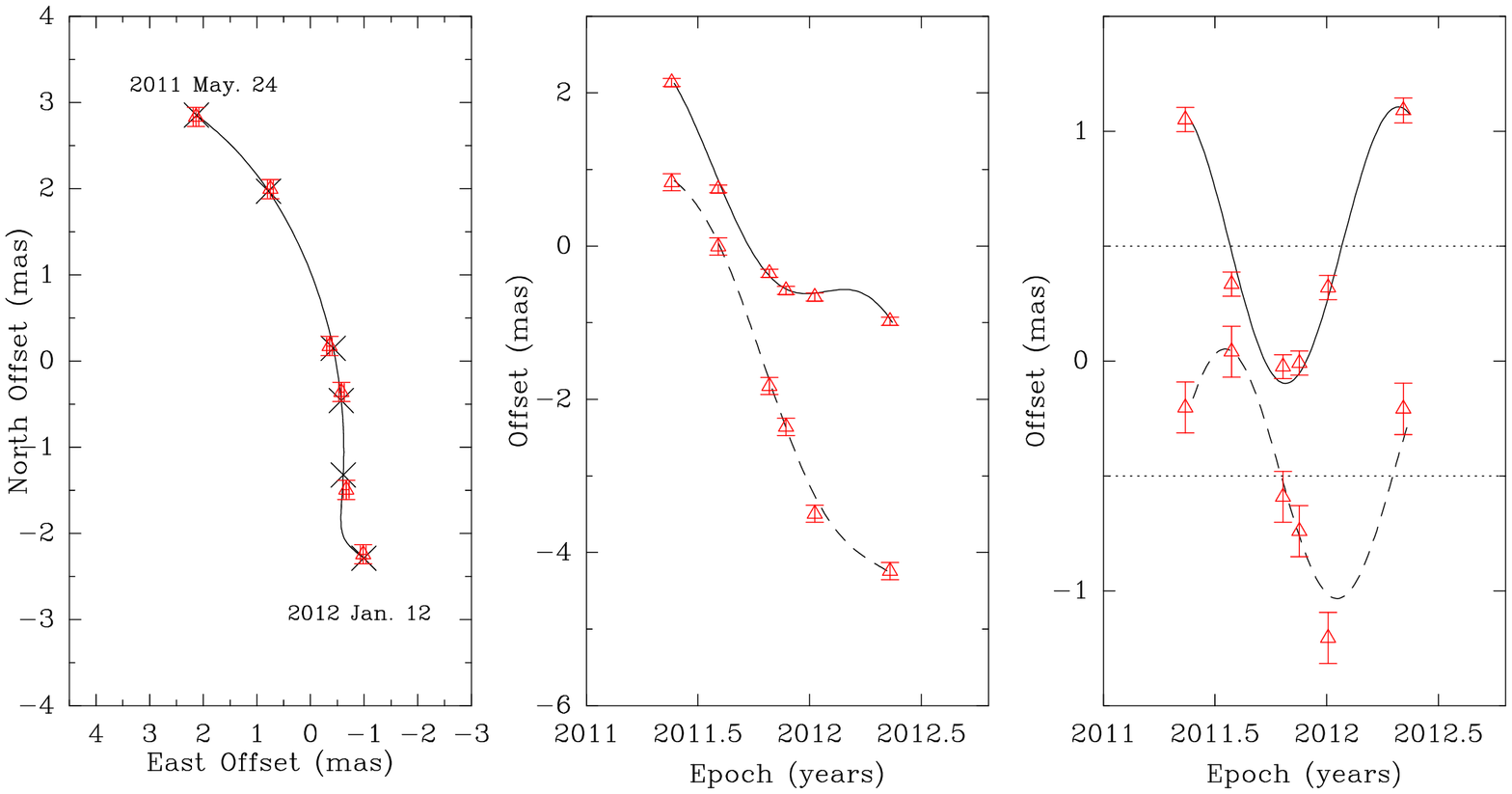}
\caption{\scriptsize Parallax and proper motion data and fits for \Gk. Plotted
are position offsets of the maser spot at \Vlsr = -4.6~\kms\ (red
triangles) relative to the background source J2007$+$4029. Solid and
dashed lines in the three panels represent the same fits as in the
corresponding panels of Fig.~\ref{on1_parallax}.
\label{g79_parallax}}
\end{figure}

\begin{figure}
\includegraphics[angle=-90,scale=0.6]{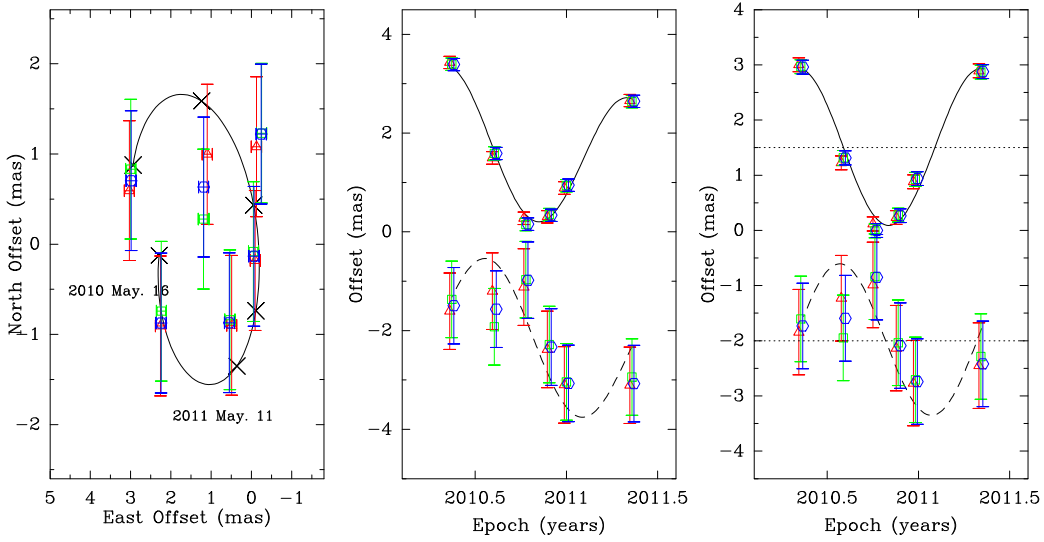}
\caption{\scriptsize Parallax and proper motion data and fits for \Gh. Plotted
are position offsets of the maser spot at \Vlsr = -6.2~\kms\
relative to three background sources J2114$+$4953 (red triangles),
J2056$+$4940 (green squares), and J2059$+$4851 (blue hexagons),
respectively. Solid and dashed lines in the three panels represent
the same fits as in the corresponding panels of
Fig.~\ref{on1_parallax}. \label{g90_parallax}}
\end{figure}

\begin{figure}
\includegraphics[angle=-90,scale=0.6]{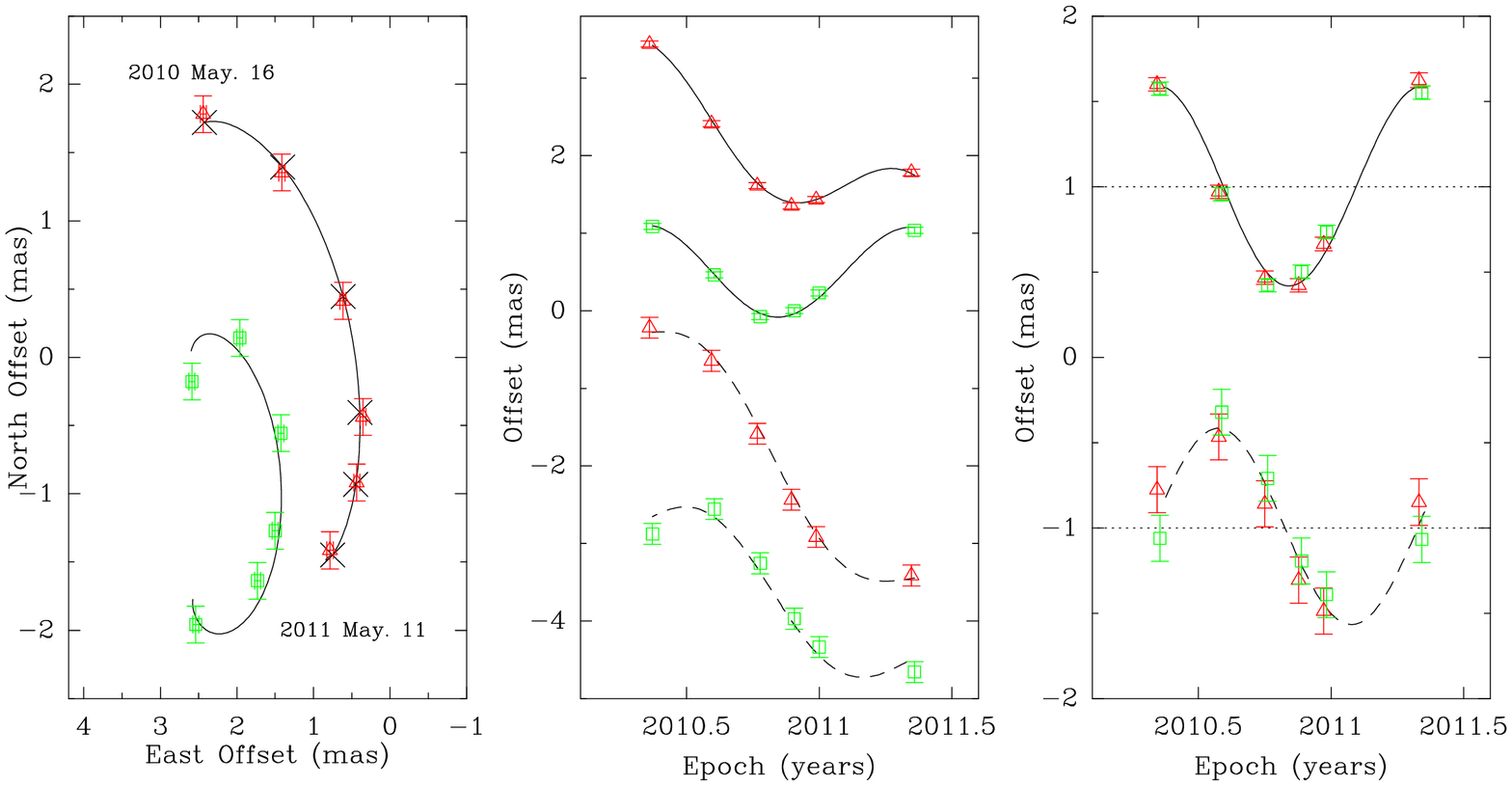}
\caption{\scriptsize Parallax and proper motion data and fits for \Gi. Plotted
are position offsets of the maser spot at \Vlsr = -3.7 (red
triangles) and -14.3~\kms\ (green square) relative to the background
sources J2117$+$5431. Solid and dashed lines in the three panels
represent the same fits as in the corresponding panels of
Fig.~\ref{on1_parallax}. \label{g92_parallax}}
\end{figure}

\begin{figure}
\includegraphics[angle=-90,scale=0.6]{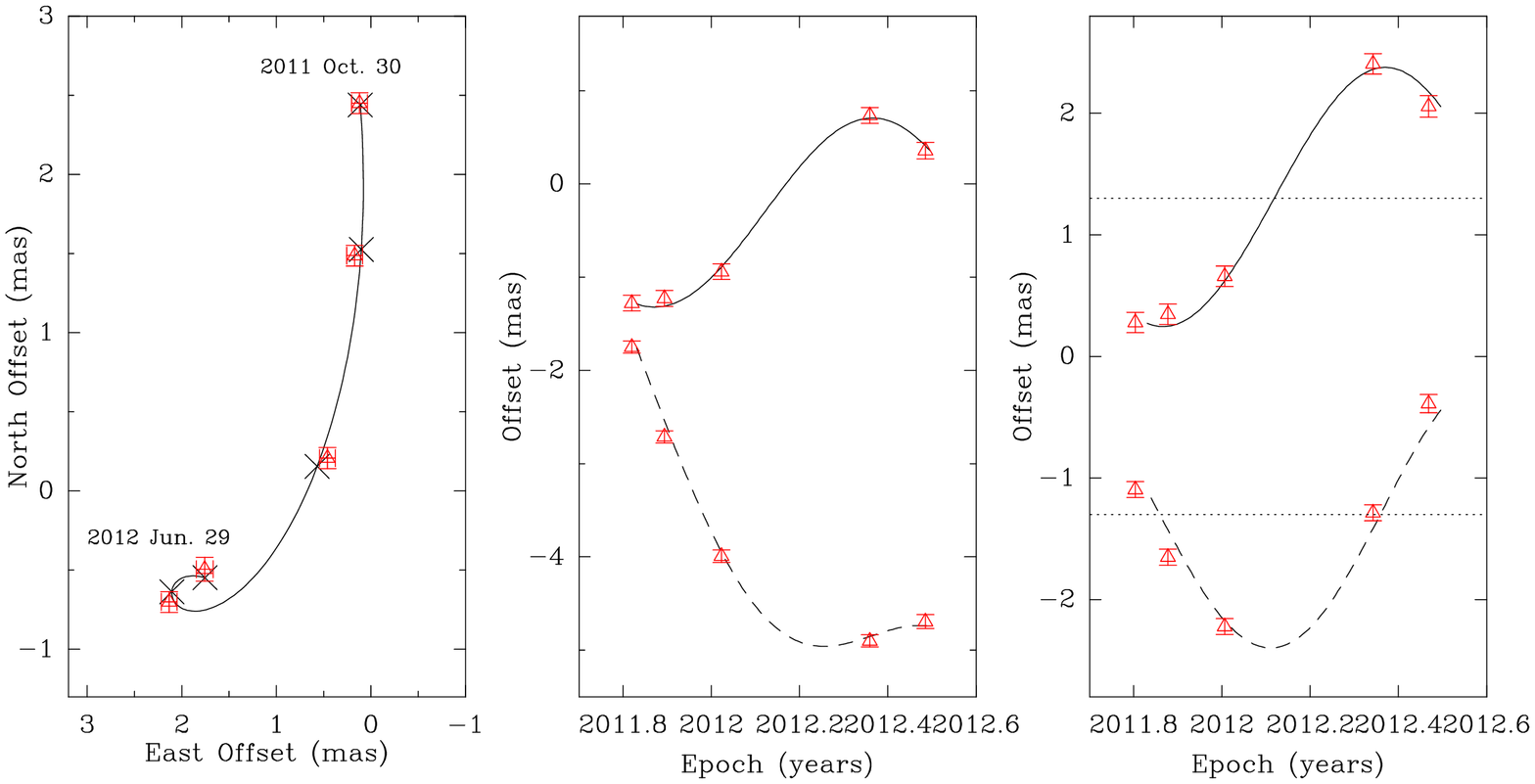}\\
\caption{\scriptsize Parallax and proper motion data and fits for \Gl. Plotted
are position offsets of the maser spot at \Vlsr = -12.1~\kms\ (red
triangles) relative to the background source J2203$+$6750. Solid and
dashed lines in the three panels represent the same fits as in the
corresponding panels of Fig.~\ref{on1_parallax}.
\label{g105_parallax}}
\end{figure}

Except for \Gh,\ we used extragalactic continuum sources
as the phase reference for the \hho\ maser sources. For \Gh\ the maser was used as the phase reference,
because the relevant three extragalactic continuum sources were too
faint to use as phase references. The quoted parallax uncertainty
is the formal fitting uncertainty, multiplied by $\sqrt{N}$ (where $N$
is the number of maser spots used in the parallax fit) to account for
possible correlations among the position measurements for the maser spots.
The fitting results are presented in Table~\ref{table:parallax}.
Figures~\ref{on1_parallax} -- \ref{g105_parallax}
show positions for the maser spots (relative to the background
sources) as a function of time and the parallax fits.

\hho\ masers can move fast (typically tens, but sometimes over 100
\kms) and spots are not always distributed uniformly around the exciting
star. This can limit the accuracy of the estimate of the proper motion of the central, exciting
star, used to study 3-D motions about the Milky Way. For \Gg,
which exhibits a distribution characteristic of a fast bipolar outflow,
its motion was determined by averaging the values from
the redshifted and blueshifted spots separately, and then averaging these two results.
With an outflow speed near 60~\kms, modest asymmetries in the outflows
could result in an uncertain average. So we assigned a proper motion uncertainty
of 20 \kms, which at its measured distance corresponds to
3~mas~yr$^{-1}$. For sources with few spots (\Gk,\ \Gh\ and \Gl),
the proper motion of reference spot was used. These sources have simple maser spectra
which span $\approx10$ \kms\ or less, suggestive of modest outflow speeds,
and we inflated the formal proper motion uncertainties to account for an unknown motion
of a single maser spot relative to the central star of 10~\kms\ at the measured
distance. For sources with a symmetric distribution of maser spots
or more than three spots that have proper motions, the motion of the central
star was determined by averaging the motion of all maser spots, such
as for ON1, \Ge,\ \Gf,\ \Gm\ and \Gi.

\begin{deluxetable}{llllllll}
\tabletypesize{\scriptsize} \tablecolumns{6} \tablewidth{0pc}
\tablecaption{Positions and Brightnesses} \tablehead{
\colhead{Source} & \colhead{R.A. (J2000)} & \colhead{Dec. (J2000)} &
\colhead{$\phi$} & \colhead{Brightness} & \colhead{\Vlsr}
& \colhead{Beam} \\
 \colhead{} & \colhead{$\mathrm{(^h\;\;\;^m\;\;\;^s)}$}
& \colhead{$(\degr\;\;\;\arcmin\;\;\;\arcsec)$} &
\colhead{($^{\circ}$)} & \colhead{(Jy/beam)} & \colhead{(\kms)}&
\colhead{(mas, mas, deg)} } \startdata
 ON 1         & 20~10~09.2036    & $+$31~31~36.090    &      & 34.1
& $+$14.8 & 0.9$\times$0.4 @ $-$13 \\
 J2003+3034   & 20~03~30.244061  & $+$30~34~30.78878  & 1.7  &  0.1
&         & 0.9$\times$0.4 @ $-$15 \\
 \\
 \Ge          & 20~25~07.1053    & $+$34~49~57.593    &      & 0.6
& $+$13.4 &  0.9$\times$0.3 @ $-$13                \\
 J2025+3343   & 20~25~10.842099  & $+$33~43~00.21448  & 1.1  & 3.9
&         &  0.9$\times$0.3 @ $-$14 \\
 \\
 \Gf          & 20~21~41.0862    & $+$37~25~29.276    &      & 5.3
& $-$9.6  &  0.8$\times$0.3 @ $-$14                \\
 J2015+3710   & 20~15~28.729794  & $+$37~10~59.51475  & 1.3  & 1.7
&         &  0.9$\times$0.3 @ $-$16 \\
 \\
 \Gm          & 20~21~44.0097    & $+$37~26~37.446    &      &58.8
& $+$3.4  &  0.7$\times$0.4 @ $-$13         \\
 J2015+3710   & 20~15~28.729794  & $+$37~10~59.51475  & 1.3  & 2.3
&         &  0.6$\times$0.4 @ $-$15 \\
 \\
 \Gg          & 20~27~25.4816    & $+$37~22~48.482    &      & 0.2
& $+$6.9 &  0.8$\times$0.3 @ $-$20  \\
 J2015+3710   & 20~15~28.729794  & $+$37~10~59.51475  & 2.4  & 1.6
&     &  0.8$\times$0.3 @ $-$22  \\
 \\
 \Gk          & 20~30~29.1464    & $+$41~15~53.590    &      & 8.0
& $-$4.6 & 0.7$\times$0.3 @ $-$19   \\
 J2007+4029   & 20~07~44.9448    & $+$40~29~48.604    & 4.4  & 1.6
&        & 0.7$\times$0.3 @ $-$20   \\
 \\
 \Gh          & 21~02~22.7007    & $+$50~03~08.309    &      & 17.0
& $-$6.2 &  0.7$\times$0.3 @ $-$9 \\
 J2056+4940   & 20~56~42.73988   & $+$49~40~06.6011   & 1.0  & 0.03
&  & 0.8$\times$0.4 @ $-$10   \\
 J2059+4851   & 20~59~57.87249   & $+$48~51~12.7002   & 1.3  & 0.05
&  & 0.8$\times$0.3 @ $-$9   \\
 J2114+4953   & 21~14~15.03646   & $+$49~53~40.9552   & 1.9  & 0.06
&  & 0.8$\times$0.3 @ $-$8   \\
 \\
 \Gi          & 21~09~21.7232    & $+$52~22~37.083    &      & 26.1
& $-$3.7 &  0.7$\times$0.3 @ $-$9 \\
 J2117+5431   & 21~17~56.484452  & $+$54~31~32.50230  & 2.5  &  0.2
&  & 0.8$\times$0.4 @ $-$4   \\
 \\
 \Gl          & 21~43~06.4628    & $+$66~06~55.183    &      &  2.2
& $-$12.1 & 0.7$\times$0.3 @ $-$10 \\
 J2203+6750   & 22~03~12.62260   & $+$67~50~47.6730   & 2.6  &  0.2
&         & 0.7$\times$0.4 @ $-$6 \\
\enddata
\tablecomments {$\phi$ is the angular separation between the maser
and the calibrator. The absolute maser position, peak brightness,
local standard of rest (LSR) velocity of the brightest spot, and
a representative 2-dimensional size and position angle of
the naturally weighted beam are listed for the first epochs for
all sources except ON 1 (2009 May 7, the second epoch) and \Gl\
(2011 October 30, the third epoch).  Position angle is defined
as east of north.} \label{table:positions}
\end{deluxetable}

\begin{deluxetable}{lllllll}
\tablecolumns{5} \tablewidth{0pc} \tablecaption{Parallaxes \& Proper
Motions} \tablehead {
 \colhead{Maser} & \colhead{$\Pi$} & \colhead{$D_{\Pi}$} &
  \colhead{$\mu_x$} &
  \colhead{$\mu_y$}
\\
  \colhead{Name}      &  \colhead{(mas)}      & \colhead{(kpc)} &
  \colhead{(\masy)} &
  \colhead{(\masy)}
            }
\startdata
ON 1& 0.425$\pm$0.036 & $2.35^{+0.22}_{-0.18}$ &$-$3.22$\pm$0.05
& $-$5.54$\pm$0.06 \\
\Ge & 0.629$\pm$0.017 & $1.59^{+0.04}_{-0.04}$ &$-$3.79$\pm$0.18
& $-$4.88$\pm$0.25 \\
\Gf & 0.285$\pm$0.022 & $3.51^{+0.29}_{-0.25}$ &$-$3.08$\pm$0.06
& $-$4.56$\pm$0.08 \\
\Gm & 0.281$\pm$0.034 & $3.56^{+0.49}_{-0.38}$ &$-$2.79$\pm$0.07
& $-$4.72$\pm$0.07 \\
\Gg & 0.770$\pm$0.053 & $1.30^{+0.10}_{-0.08}$ &$-$3.73$\pm$3.00
& $-$3.84$\pm$3.00 \\
\Gk & 0.620$\pm$0.027 & $1.61^{+0.07}_{-0.07}$ &$-$3.23$\pm$1.31
& $-$5.19$\pm$1.31 \\
\Gh & 1.483$\pm$0.038 & $0.67^{+0.02}_{-0.02}$ &$-$0.67$\pm$3.13
& $-$0.90$\pm$3.13 \\
\Gi & 0.613$\pm$0.020 & $1.63^{+0.06}_{-0.05}$ &$-$0.69$\pm$0.26
& $-$2.25$\pm$0.33 \\
\Gl & 1.129$\pm$0.063 & $0.89^{+0.05}_{-0.05}$ &$-$0.21$\pm$2.38
& $-$5.49$\pm$2.38 \\
\enddata
\tablecomments{\scriptsize Column 2 is the measured parallax. Column 3 is
the parallax converted to distance. Columns 4 and 5 are proper motion in the
eastward ($\mu_x=\ura \cos{\delta}$) and northward directions
($\mu_y=\udec$), respectively.}
\label{table:parallax}
\end{deluxetable}

\subsection{\it{Individual Sources}}
\noindent
\textbf{ON 1} is associated with IRAS 20081+3122, which coincides with an
ultra-compact H\textsc{II} (UC H\textsc{II}) region \citep{Kurtz:04}.
\citet{Yang:02} detected a high-velocity CO line wing,
which was later confirmed as a bipolar
outflow by both CO \citep{Xu:06b} and H$^{13}$CO$^{+}$ lines
\citep{Kumar:04}, indicating an active star-forming region.
This source has measured parallaxes by \citet{Nagayama:11}
with the VERA 22-GHz using \hho\ masers
and by \citet{Rygl:10} with the EVN using 6.7-GHz \meth\ masers. We present
our results in Table~\ref{table:parallax}. Within the uncertainties
all three measurements agree on the parallax, but there are slight
differences among the proper motions.
In order to calculate the peculiar motion of this source, we averaged
the results of the three measurements
(see Table~\ref{table:preper}).

\noindent \textbf{\Ge}, associated with IRAS~20231$+$3440,
is located within Lynds 870.
\hho\ masers were first detected by \citet{Palla:93}. \citet{Mao:02}
detected a bipolar outflow from CO lines, which is likely to be
driven by a low- or intermediate-mass young stellar object
(YSO). This is further confirmed by its low IRAS luminosity (500
$L_{\odot}$ at a distance of 1.6 kpc, see Table~\ref{table:parallax})
and non-detection
of both \meth\ and OH masers \citep{Walt:96,Slysh:97}, although it
satisfies the criteria that \citet{Wood:89a} used to identify
embedded massive stars and UC H\textsc{II} regions.

\noindent \textbf{\Gf}\ \& \textbf{\Gm}\ (\textbf{ON~2N}) in \textbf{ON~2}
are separated by $\approx$80$^{''}$ and associated
with two different UC H\textsc{II} regions in the complex
star-forming region ON 2 \citep{Wood:89b,Garay:93}.
\hho\ masers in \Gm\ are located $\sim$2$^{''}$ south of the peak
of 6 cm radio continuum emission \citep{Wood:89b}, which is spatially
coincident with peaks of 7 mm continuum emission and
NH$_{3}$ (3, 3) emission \citep{Ando:11,Carral:97,Codella:10}. The
masers in \Gf, which are associated with IRAS~20198$+$3716, have an offset
of $\approx25''$ from the nearest detected UC H\textsc{II} region.
The parallax and proper motion of \Gm\ were also measured
by \citet{Ando:11} with the VERA 22~GHz \hho\ masers.
Both measurements are consistent within the uncertainties, so the
peculiar motion estimates are estimated by combining both results
(see Table~\ref{table:preper}).

\noindent \textbf{\Gg} is associated with IRAS~20255$+$3712
and Sharpless 106 (S~106) IR. This region has been widely
studied at different wavelengths and
angular resolutions \citep[e.g.,][]{Kurtz:94,Smith:01,
Schneider:02,Schneider:07}. \citet{Kurtz:94} found that the UC H
\textsc{II} region G76.383--0.621 coincides with IRAS~20255$+$3712,
which is offset $\approx15''$ from the \hho\ maser position.
In the literature, the distance to \Gg\ ranges from 0.5 to 5.7 kpc
\citep{Eiroa:79,Maucherat:75}. Although a distance of $<$ 1 kpc is
commonly used \citep[e.g.,][]{Smith:01,Schneider:02},
\citet{Schneider:07} suggested that S~106 is associated with
the Cygnux X complex
at about 1.7 kpc. This parallax distance of $1.30^{+0.10}_{-0.08}$~kpc
indicates that S~106 is close to the Cygnus X
complex (parallax distance of $1.40\pm0.08$~kpc, \citet{Rygl:12}).

\noindent \textbf{\Gk} is associated with IRAS~20286$+$4105
with an infrared
luminosity of more than 5000 $L_{\odot}$ at a distance of 1.6 kpc
(see Table~\ref{table:parallax}). It satisfies the IR criteria of
\citet{Wood:89a} for UC H\textsc{II} regions, although no radio centimeter-continuum
emission is detected with the VLA \citep{Molinari:98}.

\noindent \textbf{\Gh}, associated with IRAS~21007$+$4951,
is in the dark cloud Lynds 998. Its IRAS luminosity is
$\sim$30 $L_{\odot}$ at an assumed distance of 0.67 kpc (see Table~\ref{table:parallax}),
indicating a low-mass star-forming region, although it satisfies the criteria of
\citet{Wood:89a} for an UC H\textsc{II} region. This
region has not been well studied to date. However, a CO bipolar
outflow \citep{Clark:86} and SiO emission \citep{Harju:98} indicate
active star formation.

\noindent \textbf{\Gi} is coincident with the submillimeter-continuum source
JCMTSF J210921.7+522232 \citep{Di:08}.
SiO emission has also been detected in this region \citep{Harju:98}.
There are two IRAS sources, IRAS 21078+5211 and IRAS 21078+5209,
with separations of $\approx$ 74$''$ and 60$''$, respectively. Both have
infrared luminosity of $\sim10^{4}L_{\odot}$ at our measured distance
of 1.63 kpc (see Table~\ref{table:parallax}), indicating a HMSFR.
However, \Gi\ is not likely to be directly associated
with either IRAS source given the large separations.

\noindent \textbf{\Gl} is a deeply embedded YSO \citep{Weintraub:94}
associated with
the LkH$\alpha$ 234 region in the NGC 7129 molecular cloud
\citep{Trinidad:04}. The maser position coincides with the
centimeter-continuum source VLA 3B, which is interpreted in terms of
shock-induced ionization in a thermal radio jet rather than as an
UC H\textsc{II} region \citep{Trinidad:04}.
The parallax distance of $0.89^{+0.05}_{-0.05}$~kpc is consistent
with a photometric
distance of $\sim$1 to 1.25 kpc \citep{Racine:68,Shevchenko:89}.

\section{LOCATION AND PITCH ANGLE OF THE LOCAL ARM}

\begin{figure}
\includegraphics[angle=-90,scale=0.6]{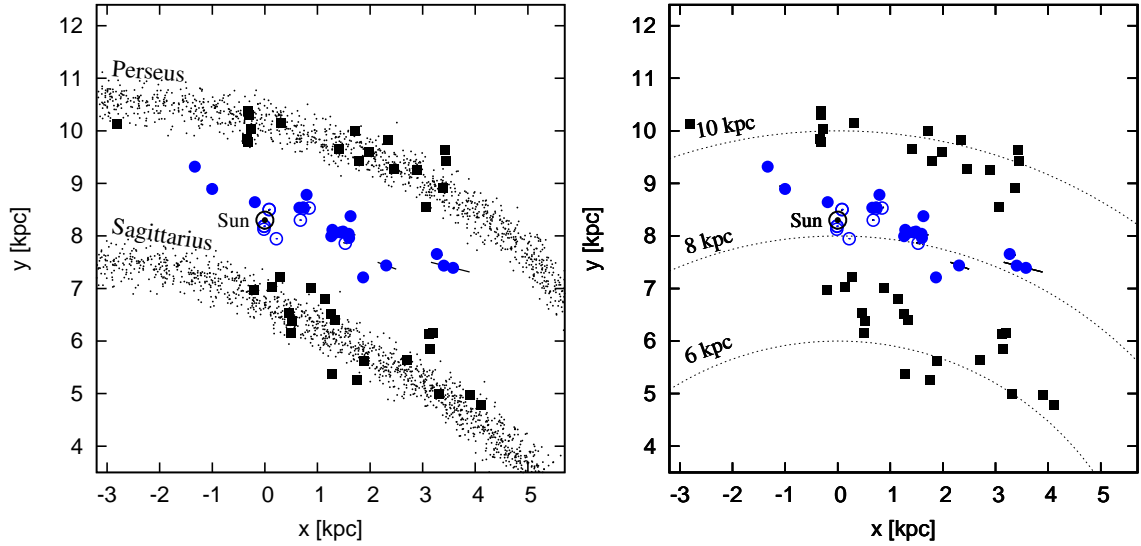}\\
\caption{\scriptsize Location of 30 sources (dots) determined by trigonometric
parallaxes in the Local arm. The filled dots represent HMSFRs,
while the unfilled ones represent low- or intermediate-mass
star-forming regions. The black squares represent some parallax data in the Perseus and Sagittarius arm.
Distance error bars of the Local arm sources are indicated, but most
are smaller than the symbols. The Galaxy is viewed from the north galactic pole with the Galactic
center (not shown) located at (0, 0) kpc. The Sun is located at (0,
8.3) kpc and marked by a black circle around a dot. The background (left panel) is a model
of the Milky Way of \citet{Cordes:02}, while the lines (right panel) are constant radius.
[See the electronic edition of the Journal for a color version of this figure.]\label{local}}
\end{figure}

Initial results from the BeSSeL Survey provide more than 60 sources
with measured parallaxes. Many of these trace the Sagittarius arm
inward from the Sun (Wu, in preparation) and the Perseus arm outward from the Sun
(Choi, in preparation), clearly locating them in the Milky Way relative to the Sun.
In addition, we find nine sources that
are between Galactocentric distances of 7.8 and 8.7~kpc toward Galactic
longitudes of $\sim70^\circ$ or $\sim105^\circ$. This places them
between the Sagittarius arm and the Perseus arm. There are another 21 sources
(including some low- and intermediate-mass star-forming regions) in the literature
that also lie between the Sagittarius and Perseus arms (see Fig.~\ref{local}).
In a pair of accompanying papers (Choi, in preparation; Wu, in preparation) we find
that the Sagittarius and Perseus arms appear to be clear structures in longitude, velocity and
distance.  Here we define members of the Local arm as all sources that lie between
these two structures, finding that they show a coherent pattern in space.
Table~\ref{table:preper} summarizes the parallax, proper motion, and \vlsr\ of these 30 sources.

The 30 star-forming regions in the Local arm trace a narrow,
slightly curved, ``swath'' starting at $\approx3.6$ kpc toward
$l\approx 60^{\circ}$ and extending $\approx1.9$ kpc toward $l \approx 230^{\circ}$
(Since there currently are no parallax distances measured between
$l = 240^{\circ}$ and $270^{\circ}$, it is not clear how far the
Local arm extends in the $3^{rd}$ Galactic quadrant.)
Visually, the width of the swath is less than about 1 kpc, while
the separation between the Sagittarius and Perseus arms in this
region of the Galaxy is about 3.5 kpc. The Sun is located close
to the inner edge of the arm.

Comparing the locations of the Sagittarius and Perseus arms, based on
parallaxes (Choi, in preparation; Wu, in preparation), the distance between the centers of Local
and Sagittarius arms of is $\approx2$ kpc and between the Local and Perseus arms
is $\approx$ 1.5 kpc (both measured along a line from the
Galactic center through the Sun). Thus, contrary to previous claims
\citep[GG76,][]{Val:02}, the Local arm is nearer to the Perseus than
to the Sagittarius arm.

It has been suggested that the Local arm is an interarm branch or
spur (GG76). Spurs and branches are often observed in external
galaxies \citep{Elm:80,Scoville:01,La:06} and
are thought to have lifetimes comparable to those
of arms \citep{Elm:80}. However, there are distinct differences
between spurs and branches. It is generally thought that spurs extend outward
from an arm into the interarm space at a large pitch angle
\citep[$\sim$60$^{\circ}$,][]{Elm:80},
whereas branches have a smaller pitch angle (usually $<$20$^{\circ}$)
and are longer than spurs. A branch usually occurs as a bifurcation
of a main spiral arm and gives rise to the main interarm features, as
seen in Figures 9 and 12 of \citet{La:06}.  Based on our observations,
the Local arm is clearly not a spur and instead may be a branch.

\begin{figure}
\includegraphics[scale=0.9]{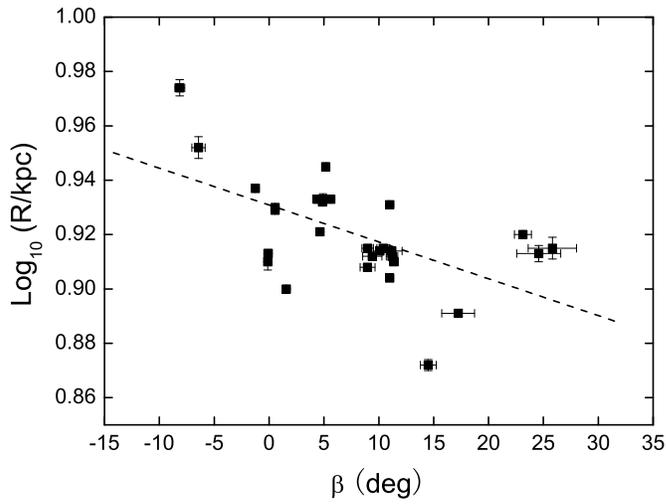}
\caption{\scriptsize Pitch angle of the Local arm. The logarithm of
Galactocentric radius R (in kpc units) is plotted against
Galactocentric azimuth $\beta$, defined as zero toward the Sun
and increasing with increasing Galactic longitude.   Data based on trigonometric
parallaxes for sources in Table~\ref{table:preper} are shown along
with 1$\sigma$ uncertainties. Positional variations of these sources
are clearly greater than the parallax uncertainties. The fit line
to the data (including all sources) is shown with dashed line. Pitch angle
is proportional to the negative of the arctangent of the line slope and
we estimate a pitch angle of $10.1^{\circ} \pm 2.7^{\circ}$.\label{angle}}
\end{figure}

We estimate the pitch angle of the Local arm by fitting a straight
line to the logarithm of Galactocentric radius, $R_{gc}$, versus Galactocentric
azimuth, $\beta$ (defined as 0 toward the Sun and
increasing with Galactic longitude), as shown in Figure~\ref{angle}. We do the fitting with a
Bayesian Markov chain Monte Carlo (McMC) exploration of parameter space using the
Metropolis-Hastings algorithm to accept or reject trials.  Since parallax
uncertainty maps into both $\ln{R_{gc}}$ and $\beta$, we numerically
evaluate uncertainties for these parameters and then minimize residuals
divided by their uncertainties projected perpendicular to the fitted line.
Since the slope of the fitted line is the tangent of the pitch angle,
one of the two fitted parameters, we iterate the fitting procedure until
convergence before performing the final McMC trials used to determine
marginalized posteriori probability density functions for the two parameters.
See Reid et al. (in preparation) for more details. We find the pitch angle of the
Local arm to be 10.1$^{\circ}\pm2.7^{\circ}$. This value is shallower than the preliminary
estimate in \citet{Reid:09b} based on only five parallaxes and heavily
weighted by G059.78+0.06, which may be an outlier. Our estimated spiral
arm pitch angle is similar to that of the Sagittarius and Perseus arms (Wu, in preparation; Choi, in preparation)
and is characteristic of major arms in Sb-Sc type galaxies \citep{Kennicutt:81}.

Our findings suggest three possibilities for the nature
of the Local arm in relation to the spiral structure of the Milky Way.

\begin{figure}
\includegraphics[angle=-90,scale=0.6]{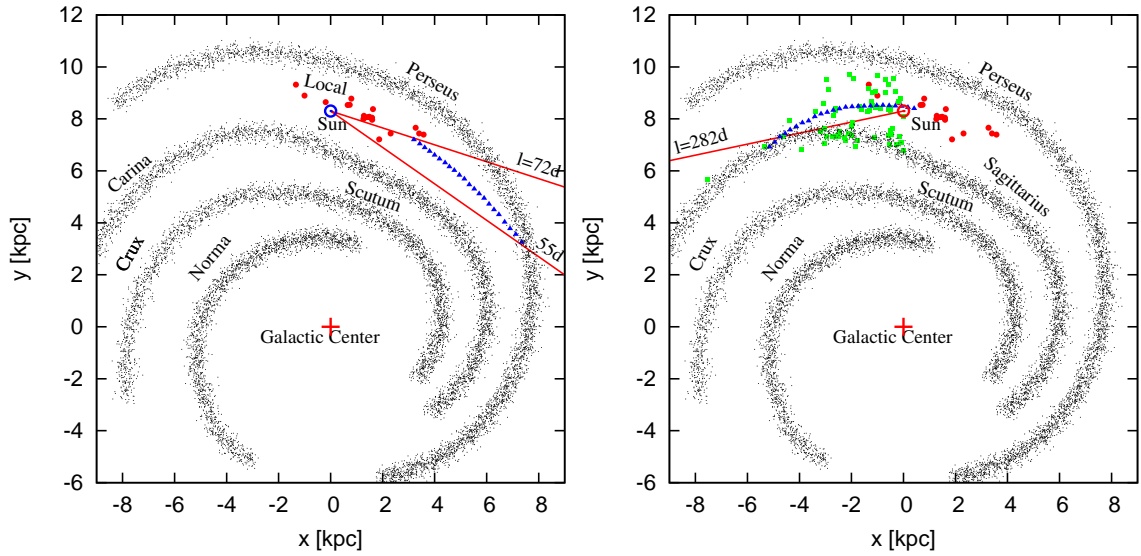}
\caption{\scriptsize Two possible models of the Local arm.
Red dots represent the parallax data in Table~\ref{table:preper}.
Blue triangles represent two possible models of the Local arm.
Green squares represent stellar distances \citep{Rus:03,Rus:07}.
The background is a model of the Milky Way of \citet{Cordes:02}.
[See the electronic edition of the Journal for a color version of this figure.]
\label{carina}}
\end{figure}

(1) It could be a branch of the Perseus arm, as suggested by many
authors.  Two facts support this theory. The Local arm is closer to the Perseus
arm than to the Sagittarius arm, and branches often appear at the inner
edge of an arm in external galaxies \citep{Patsis:97,La:06}.
Extrapolating from the most distant measured star-forming region in
Figure~\ref{carina} (left panel) suggests that a bifurcation point
could be at a distance of $\approx6$ kpc at $l \approx55^{\circ}$.
This location is close to that suggested by \citet{Ura:87} for such
a branching, based on photometric distances of open star clusters.

(2) The Local arm could be part of a major arm, as suggested by Bok and
Kerr \citep{Bok:59,Bok:70,Kerr:70,KK:70}. In their model, the Local
arm runs through the Sun and connects with the Carina arm, although most authors
following the GG76 model believe that the Carina arm connects to the Sagittarius
arm.  At present, however, the lack of parallax data for $l > 230^{\circ}$
precludes critically testing this theory.  Figure~\ref{carina} (right panel) shows
the positions of optical H \textsc{II} regions with stellar distances
in the 3rd Galactic quadrant from \citet{Rus:03}. It is possible
the Local arm connects to the Carina arm at $l\approx282^{\circ}$,
near its tangent point \citep{Bro:00}. This could be a bifurcation
point where the Carina arm splits into the Local branch and the Sagittarius
branch. However, since optical distances generally have large uncertainties,
as shown by \citet{Rus:07}, more VLBI parallax results from the Southern
Hemisphere are required before a definite conclusion can be reached.

(3) The Local arm is an independent spiral arm segment, which
consists of numerous major star-forming regions, such as Cygnus X region \citep{Rygl:12},
with similar Galactocentric distances and space velocities (see Section~\ref{sect:3-D}).
From the model of \citet{Cordes:02}, the interarm spacing between
two major arms of the Milky Way is usually more than 2 kpc, even
more than 3 kpc in some places. If the Local arm is a major arm,
the spacing between it and nearby arms would be only
1 -- 2 kpc. However, such a small spacing is not expected from the theory
of spiral density waves \citep{Yuan:69}.

\section{SPACE MOTION OF SOURCES IN THE LOCAL ARM} \label{sect:3-D}

Combining the distances, LSR velocities and proper motions of the
masers in star forming regions yields their locations in the
Galaxy and their full space motions. The proper motion uncertainties for
most sources listed in Table \ref{table:preper} are based on
measurement accuracy alone. There is
additional uncertainty when referring these motions to that
of the central star (or stars) that excite the masers. Basically, they
are 3 to 5 \kms\ for \meth,\ SiO masers and continuum emission,
and 5 to 20 \kms\ for \hho\ masers (depending on the width/complexity
of the spectrum). The \vlsr\ values are based on weighted average
velocity of maser and the larger molecular cloud material
(usually from CO or other thermally emitting molecule lines) of which
the maser is a part. For water masers, we weighted the thermal line velocity more than water
masers, while for \meth\ and SiO masers a greater weight is given to
masers, as they are much more closely tied to the central star.
The uncertainties for \vlsr\ include a term that comes from the difference
between a maser and its associated thermal line velocity. For continuum sources,
the \vlsr\ values are based on the velocity of thermal lines and we assign
uncertainties of $\pm$5 \kms.

Given a model for the scale and rotation of the Milky Way, we can subtract the
effects of Galactic rotation and estimate peculiar (i.e. non-circular) motions.
Peculiar motions are given in a Galactocentric reference frame,
where  $U_{s}, V_{s} \, {\rm and} \, W_{s}$  are the velocity components toward
the Galactic center, in the direction of Galactic rotation, and toward the North
Galactic Pole, respectively, at the location of a given source in the Galaxy.
Details of these calculations are given in the Appendix of \citet{Reid:09b}.
Here we adopt recent estimates of $R_{0} = 8.3$ kpc and
$\Theta_{0} = 239$~\kms\ \citep{Brunthale11}, and the latest Solar motion
values $U_{\odot} = 11.10$ \kms,\ $V_{\odot} = 12.24$ \kms\ and
$W_{\odot} = 7.25$ \kms\ \citep{Schonrich10}.

\begin{figure}
\includegraphics[angle=-90,scale=0.6]{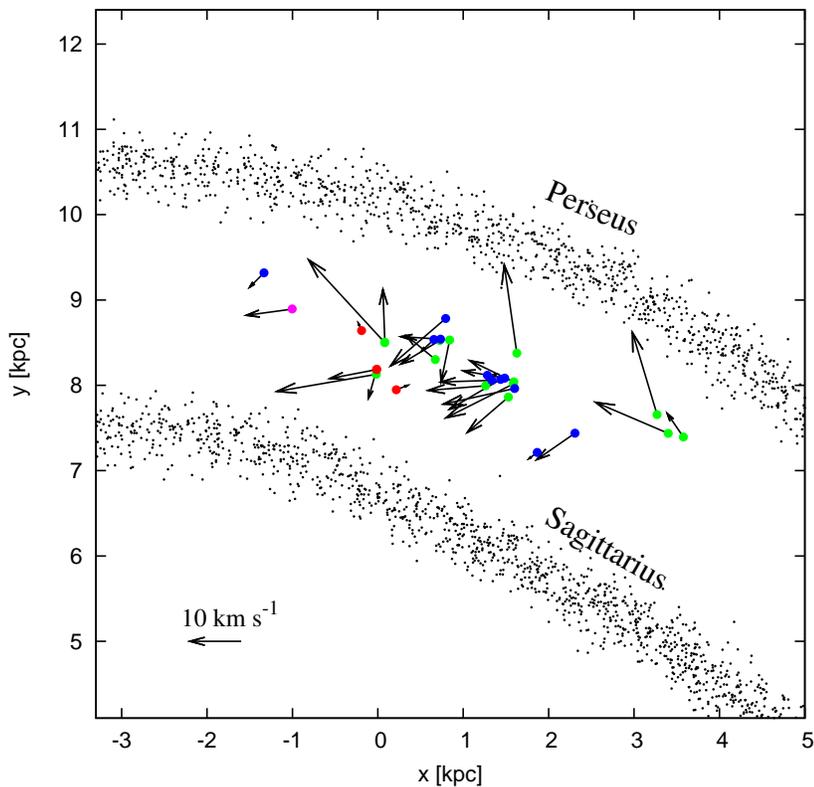}
\caption{Peculiar motion vectors of methanol (blue dots), water (green dots),
SiO (pink dot) masers, and continuum emission sources (red dots) in the Local
arm after transforming to a reference frame rotating with the galaxy,
using values of $R_{0} = 8.3$~kpc and $\Theta_{0} = 239$ \kms\ \citep{Brunthale11}.
A motion scale of 10~\kms\ is indicated in the bottom left corner of the
panels. The background is a model of the Milky Way of \citet{Cordes:02}.
The galaxy is viewed from the north galactic pole, it rotates clockwise, and the Sun
is at (0,8.3) kpc. [See the electronic edition of the
Journal for a color version of this figure.] \label{local_3D}}
\end{figure}

The peculiar motions of the 30 Local arm sources are shown in the
Figure~\ref{local_3D} and range from $-16$ to 10 \kms\ towards the Galactic center and
from $-19$ to 2 \kms\ in the direction of Galactic rotation (Table \ref{table:3d}).
A weighted average of these peculiar motions indicates a systematic peculiar motion for
Local arm sources  of $1.3\pm1.3$ \kms\ toward the Galactic center ($\overline{U_s}$),
$-5.4\pm1.0$ \kms\ in the direction of Galactic rotation ($\overline{V_s}$), indicating
counter rotation, and $2.4\pm1.4$ \kms\ toward the north Galactic pole ($\overline{W_s}$).
We use the standard error of the means as the uncertainties.
Note that the values for ${U_s}$ and ${V_s}$ depend sensitively on the adopted
$\Theta_{0}$ and $V_{\odot}$ values, respectively.
From a sample of 16 HMSFRs in several spiral arms, \citet{Reid:09b} obtained
($\overline{U_s}$, $\overline{V_s}$, $\overline{W_s}$) = ($2, -15, -3$) \kms\
with uncertainties of $\pm$2 \kms.  These average peculiar motions used the Solar Motion
from \citet{Dehnen:98} which had $V_{\odot} = 5.2$ \kms, or about 7 \kms\ lower than
the \citet{Schonrich10} value. Adjusting the \citet{Reid:09b} values to the newer
Solar Motion would yield ($\overline{U_s}$, $\overline{V_s}$, $\overline{W_s}$)
$\approx(2, -8, -3)$ \kms. Thus, the Local arm peculiar motions, on average, are consistent with those of
HMSFRs in other spiral arms, further supporting our theory
that the Local arm is a major structure in the Milky Way.

\section{CONCLUSIONS}

We have studied the nature of the Local arm by measuring parallax
distances and proper motions of nine 22-GHz \hho\ masers associated
with star-forming regions. We include previously published
parallaxes and proper motions for 21 other star forming regions
(from either \hho,\ \meth,\ SiO masers or YSO continuum emission).
These 30 sources clearly lie between the Sagittarius and Perseus
spiral arms and belong to the Local arm. This arm is at least
$\sim$5 kpc in length and $\sim$1 kpc in width; it is not a spur and
may be a branch of the Perseus arm, a bifurcation of the Carina arm,
or an independent arm segment. The average peculiar motions
of the sources in the Local arm are similar to those of HMSFRs in other
major spiral arms of the Milky Way.

\vskip 0.5truecm {\it Facilities:} \facility{VLBA}

\acknowledgments We would like to thank the anonymous referee for
many useful comments that have improved the paper. This work was supported by
the Chinese NSF through grants NSF 11133008, NSF 11073054, NSF 11203082, BK2012494, and the Key
Laboratory for Radio Astronomy, CAS. This research has made use of the SIMBAD database,
operated at CDS, Strasbourg, France.


\begin{deluxetable}{lrrrrrrc}
\tabletypesize{\scriptsize} \tablewidth{0pt} \tablecolumns{8}
\tablecaption{Parallaxes and Proper Motions for the Local arm}
\tablehead {
  \colhead{Source} & \colhead{$\ell$} & \colhead{$b$} &
  \colhead{Parallax} & \colhead{$\mu_x$} & \colhead{$\mu_y$} &
  \colhead{\vlsr}    & \colhead{Ref.}
\\
  \colhead{}      & \colhead{(deg)} & \colhead{(deg)} &
  \colhead{(mas)} & \colhead{(\masy)} & \colhead{(\masy)} &
  \colhead{(\kms)}& \colhead{}
           }
\startdata

EC95/Serpens                    & 31.5615 &$+$5.3303 &$2.410\pm0.020$ &$+0.70\pm0.02$ &$-3.64\pm0.10$ &$7\pm5$ & 5,8 \\
\Gsrc                           & 59.7829 &$+$0.0647 &$0.463\pm0.020$ &$-1.65\pm0.03$ &$-5.12\pm0.08$ &$25\pm3$ & 25,32 \\
ON 1\tablenotemark{a}           & 69.5402 &$-$0.9755 &$0.406\pm0.031$ &$-3.19\pm0.37$ &$-5.22\pm0.25$ &$12\pm5$ & 1,24,25,27 \\
\Ge                             & 74.0343 &$-$1.7140 &$0.629\pm0.017$ &$-3.79\pm0.18$ &$-4.88\pm0.25$ &$5\pm5$ & 1,4 \\
\Gf                             & 75.7610 &$+$0.3400 &$0.285\pm0.022$ &$-3.08\pm0.06$ &$-4.56\pm0.08$ &$-9\pm9$ & 1,25 \\
\Gm/ON 2N\tablenotemark{b}      & 75.7821 &$+$0.3428 &$0.271\pm0.022$ &$-2.79\pm0.10$ &$-4.69\pm0.12$ &$1\pm5$ & 1,2,25 \\
\Gg                             & 76.3809 &$-$0.6177 &$0.770\pm0.053$ &$-3.73\pm3.00$ &$-3.84\pm3.00$ &$-2\pm5$ & 1,4 \\
IRAS 20126+4104\tablenotemark{c}& 78.1224 &$+$3.6328 &$0.610\pm0.020$ &$-4.14\pm0.13$ &$-4.14\pm0.13$ &$-4\pm5$ & 4,22 \\
AFGL 2591                       & 78.8867 &$+$0.7090 &$0.300\pm0.010$ &$-1.20\pm0.32$ &$-4.80\pm0.12$ &$-6\pm7$ & 25,28 \\
IRAS 20290+4052                 & 79.7358 &$+$0.9905 &$0.737\pm0.062$ &$-2.84\pm0.09$ &$-4.14\pm0.54$ &$-3\pm5$ & 4,28 \\
\Gk                             & 79.8769 &$+$1.1770 &$0.620\pm0.027$ &$-3.23\pm1.31$ &$-5.19\pm1.31$ &$-5\pm10$ & 1,4 \\
NML~Cyg                         & 80.7984 &$-$1.9209 &$0.620\pm0.047$ &$-1.55\pm0.42$ &$-4.59\pm0.41$ &$-3\pm3$ & 15,34 \\
DR 20                           & 80.8615 &$+$0.3834 &$0.687\pm0.038$ &$-3.29\pm0.13$ &$-4.83\pm0.26$ &$-3\pm5$ & 4,28 \\
DR 21                           & 81.7524 &$+$0.5908 &$0.666\pm0.035$ &$-2.84\pm0.15$ &$-3.80\pm0.22$ &$-3\pm3$ & 7,28\\
W 75N                           & 81.8713 &$+$0.7807 &$0.772\pm0.042$ &$-1.97\pm0.10$ &$-4.16\pm0.15$ &$7\pm3$ & 25,28 \\
\Gh                             & 90.2105 &$+$2.3244 &$1.483\pm0.038$ &$-0.67\pm3.13$ &$-0.90\pm3.13$ &$-3\pm5$ & 1,4 \\
\Gi                             & 92.6712 &$+$3.0712 &$0.613\pm0.020$ &$-0.69\pm0.26$ &$-2.25\pm0.33$ &$-5\pm10$ & 1,4 \\
\Gl                             & 105.4154 &$+$9.8783 &$1.129\pm0.063$ &$-0.21\pm2.38$ &$-5.49\pm2.38$ &$-10\pm5$  & 1,25 \\
IRAS 22198+6336                 & 107.2982 &$+$5.6398 &$1.309\pm0.047$ &$-2.47\pm0.21$ &$+0.26\pm0.40$ &$-11\pm5$ & 4,10,13 \\
L 1206                          & 108.1845 &$+$5.5188 &$1.289\pm0.153$ &$+0.27\pm0.23$ &$-1.40\pm1.95$ &$-11\pm3$  & 20,27 \\
Cep~A                           & 109.8713 &$+$2.1143 &$1.430\pm0.080$ &$+0.50\pm1.10$ &$-3.70\pm0.20$ &$-7\pm5$  & 21,25 \\
L 1287                          & 121.2978 &$+$0.6588 &$1.077\pm0.039$ &$-0.86\pm0.11$ &$-2.29\pm0.56$ &$-23\pm5$  & 25,27 \\
L~1448~C                        & 158.0625 &$-$21.4203 &$4.310\pm0.330$ &$+21.90\pm0.70$ &$-23.10\pm3.30$ &$10\pm7$ & 3,12 \\
SVS 13/NGC 1333                 & 158.3474 &$-$20.5551 &$4.250\pm0.320$ &$+14.25\pm1.30$ &$-8.95\pm1.75$ &$7\pm3$ & 11,29 \\
Orion\tablenotemark{d}          & 209.0071 &$-$19.3854 &$2.408\pm0.033$ &$+3.30\pm1.50$ &$+0.10\pm1.50$ &$3\pm5$ & 9,16,19,31 \\
G232.62$+$00.99                 & 232.6205 &$+$0.9956 &$0.596\pm0.035$ &$-2.17\pm0.06$ &$+2.09\pm0.46$ &$21\pm3$ & 4,26 \\
VY~CMa\tablenotemark{e}         & 239.3526 &$-$5.0655 &$0.855\pm0.080$ &$-2.80\pm0.20$ &$+2.60\pm0.20$ &$20\pm3$ & 6,23,33\\
DoAr21/Ophiuchus                & 353.0157 &$+$16.9815 &$8.200\pm0.370$ &$-26.47\pm0.92$ &$-28.23\pm0.73$ &$3\pm5$ & 17,18 \\
S1/Ophiuchus                    & 353.0992 &$+$16.8943 &$8.550\pm0.500$ &$-3.88\pm0.87$ &$-31.55\pm0.69$ &$3\pm5$ & 17,18 \\
IRAS 16293-2422                 & 353.9354 &$+$15.8393 &$5.600\pm1.500$ &$-20.60\pm0.70$ &$-32.40\pm2.00$ &$4\pm3$ & 14,30 \\

\enddata
\tablecomments{\scriptsize Columns 2 and 3 give Galactic longitude
and latitude, respectively. Columns 5 and 6 are proper motion in the
eastward ($\mu_x=\ura \cos{\delta}$) and northward directions
($\mu_y=\udec$), respectively. Column 7 lists LSR velocity
components of molecular line emission; these can be converted to a
heliocentric frame as described in the Appendix of
\citet{Reid:09b}.} \tablecomments{References are
1: this paper; 2: \citet{Ando:11}; 3: \citet{Bachiller:90} CS (1-0);
4: \citet{Bro:96} CS (2-1); 5: \citet{Choi:99} HCO$^{+}$ (1-0); 6: \citet{Choi:08};
7: \citet{Dic:78} CO (1-0); 8: \citet{Dzib:10}; 9: \citet{Hirota:07};
10: \citet{Hirota:08a}; 11: \citet{Hirota:08b}; 12: \citet{Hirota:11};
13: \citet{Honma:12}; 14: \citet{Imai:07}; 15: \citet{Kemper:03} CO (4-3);
16: \citet{Kim:08}; 17: \citet{Loinard:08}; 18: \citet{Loren:89} $^{13}$CO (1-0);
19: \citet{Menten:07}; 20: \citet{Mol:96} NH$_{3}$ (1,1)(2,2);
21: \citet{Mos:09}; 22: \citet{Mos:11}; 23: \citet{Muller:07} CO (2-1);
24: \citet{Nagayama:11}; 25: \citet{Plume:92} CS (7-6); 26: \citet{Reid:09a};
27: \citet{Rygl:10}; 28: \citet{Rygl:12}; 29: \citet{Snell:81} CO (1-0);
30: \citet{Tak:07} HCN (4-3); 31: \citet{Wiseman:98} NH$_{3}$ (1,1);
32: \citet{Xu:09}; 33: \citet{Zhang:12a}; 34: \citet{Zhang:12b}}
\tablenotetext{a}{The parallax and proper
motions of ON 1 are unweighted average values of the independent
results from \citet{Rygl:10}, \citet{Nagayama:11} and this paper.}
\tablenotetext{b}{The parallax and proper motions of \Gm\ is unweighted
average value
of the independent results from \citet{Ando:11} and this paper.}
\tablenotetext{c}{The proper motion of IRAS~20126+4104 is from its
methanol masers (Moscadelli, private communicaiton).}
\tablenotetext{d}{The parallax of Orion is an unweighted average value
of the independent results from \citet{Menten:07} and
\citet{Kim:08} with equal accuracy, while the proper motions come from
\citet{Menten:07}, because \citet{Hirota:07} and \citet{Kim:08} only
use one spot to estimate these values.}
\tablenotetext{e}{The parallax of VY~CMa is unweighted average value
of the independent results from \citet{Choi:08} and
\citet{Zhang:12a}, while the proper motions come from
\citet{Zhang:12a} because \citet{Choi:08} only use one spot to estimate
these values.}\label{table:preper}
\end{deluxetable}

\begin{deluxetable}{lrrrc}
\tabletypesize{\scriptsize} \tablewidth{0pc} \tablecaption{Peculiar
Motion of Sources in the Local arm} \tablehead {
  \colhead{Source} & \colhead{U$_{s}$} & \colhead{V$_{s}$} &
  \colhead{W$_{s}$} & \colhead{Emission}
\\
  \colhead{}      & \colhead{(\kms)} & \colhead{(\kms)} &
  \colhead{(\kms)}& \colhead{}
           }
\startdata

EC95/Serpens & $-0.9\pm4.4$ & $ 2.2\pm3.5$   & $ 1.8\pm0.8 $  & C \\
\Gsrc        & $1.5\pm3.1$  & $-1.2\pm3.6$   & $-4.2\pm0.9 $  & M \\
ON1          & $7.1\pm4.7$  & $-5.6\pm5.5$   & $ 5.3\pm4.0 $  & M,W \\
\Ge          & $8.3\pm2.8$  & $-6.5\pm5.4$   & $ 9.7\pm1.7 $  & W \\
\Gf          & $0.4\pm4.7$  & $-15.5\pm9.2$  & $ 6.0\pm1.3 $  & W \\
\Gm/ON 2N    & $-2.8\pm5.1$ & $-4.9\pm5.6$ & $0.5\pm2.1$ & W \\
\Gg          & $2.7\pm18.6$  & $-10.8\pm5.6$  & $12.4\pm18.6$   & W \\
IRAS 20126+4104 & $5.7\pm2.5$  & $-13.0\pm5.4$  &  $14.9\pm1.3$ & M,W \\
AFGL 2591  & $-12.8\pm5.4$  & $-10.7\pm7.1$  & $-22.2\pm4.4$  & W \\
IRAS 20290+4052 & $1.9\pm3.5$  &  $-9.8\pm5.5$ &   $6.0\pm2.2$  & M \\
\Gk           &   $9.1\pm10.2$  &  $-11.5\pm10.2$ &   $3.4\pm10.2$  & W \\
NML~Cyg       &  $-2.4\pm4.0$  &  $-9.1\pm3.7$ &   $-4.8\pm3.4$  & W \\
DR 20         &   $7.5\pm2.6$  &  $-8.8\pm5.4$  &   $5.1\pm1.5$  & M \\
DR 21         &   $0.0\pm2.6$  &  $-8.3\pm3.6$  &   $6.6\pm1.4$  & M \\
W 75N         &  $-0.5\pm2.2$  &  $1.7\pm3.7$   &   $1.2\pm1.0$  & M \\
\Gh           &  $-4.1\pm10.1$  &  $-5.9\pm5.5$  &   $6.2\pm10.0$  & W \\
\Gi           &  $-16.4\pm4.0$ &  $-5.6\pm10.0$  & $-1.7\pm2.4$   & W \\
\Gl           & $8.4\pm9.7$   & $-0.8\pm6.4$   & $-13.2\pm10.0$   & W \\
IRAS 22198+6336 &$5.2\pm2.6$   & $-7.1\pm5.1$   & $10.6\pm1.5$  & W \\
L1206         & $0.1\pm4.4$  & $-7.7\pm4.0$   & $ 0.1\pm6.3$  & M \\
Cep~A         & $ 2.5\pm4.0$  & $-2.3\pm5.2$   & $-5.2\pm1.9$  & M \\
L1287         & $10.2\pm3.4$  & $-9.9\pm4.6$   & $-3.0\pm2.6$  & M \\
L~1448~C      &$-16.0\pm6.2$  &$-15.0\pm4.8$   & $-4.4\pm3.8$  & W \\
SVS 13/NGC 1333 &$-10.2\pm3.0$  & $-0.4\pm3.2$   & $ 4.1\pm2.1$ & W \\
Orion        & $-1.4\pm4.7$  & $-0.7\pm4.1$   & $ 5.8\pm3.3$ & C \\
G232.62$+$00.99 & $ 2.4\pm3.6$  & $-3.1\pm3.9$   & $ 0.8\pm2.0$ & M \\
VY~CMa       & $ 0.2\pm2.7$  & $-9.1\pm3.4$   & $-3.2\pm1.5$ & S\\
DoAr21/Ophiuchus & $ 1.7\pm4.9$  & $-9.1\pm2.5$   & $ 5.8\pm1.7$ & C \\
S1/Ophiuchus     & $ 5.7\pm4.9$  & $-1.7\pm2.4$   & $-4.4\pm1.7$ & C \\
IRAS 16293-2422  & $ 3.3\pm3.3$  &$-19.2\pm15.4$  & $ 1.1\pm2.7$ & W \\
\enddata
\tablecomments {{\footnotesize} C, M, W, and S denote peculiar motion
derived from continuum emission of YSOs, methanol masers, water masers, and
silicon monoxide masers, respectively.
These motions assume $R_{0} = 8.3$ kpc and
$\Theta_{0} = 239$~\kms\ \citep{Brunthale11}, and Solar Motion
values $U_{\odot} = 11.10$ \kms\, $V_{\odot} = 12.24$ \kms\, and
$W_{\odot} = 7.25$ \kms\ \citep{Schonrich10} } \label{table:3d}
\end{deluxetable}

\clearpage
\section{Online Material}\label{online}

\subsection{Parallax and Proper Motion Fitting Details}

Parallax fits are based on features persisting over 4 or more epochs.
Internal maser motions are measured for features persisting over at least 3
epochs. Table~\ref{table:detail} presents the results. The uncertainties
of parallaxes and proper motions given in this table
are the formal fitting uncertainties.

\Ge: We detected 13 features persisting over at least 3 epochs,
only 3 of which lasted 6 epochs. These features mainly consist of
three small groups located at (0,0), (0\farcs1,--0\farcs05) and
(0\farcs05,0\farcs3), as shown in Fig.~\ref{g74_rel_pro}. The three
features persisting over 6 epochs, used for the parallax fit, are
located in the group at (0,0). We average the absolute proper
motions of three groups as the source's proper motion.

\Gf: There are 10 features persisting over at least 3 epochs, 6 of
which lasted 6 epochs. All features are distributed in a ring and
expand symmetrically relative the geometric center of the six persistent
features (see Fig.~\ref{g75_rel_pro}). We average all the measured absolute proper
motions as the source's proper motion.

\Gm: We detect 20 features persisting over at least 3 epochs,
5 of which lasted 4 epochs. Fig~\ref{g75.78_rel_pro} shows the proper
motion relative to the center of the maser spots. We average all the measured
absolute proper motions as the source's proper motion.

\Gg: We detect three features persisting at least 3 epochs,
belonging to two groups separated by about 80 mas in the NE-SW
direction. A single feature lasted for 4 epochs and it is used for
the parallax fit. The LSR velocities of the NE group are
red shifted with respect to the SW group.
The blue-shifted features to the SW expand with
a very high velocity ($\sim$143~\kms) relative to the redshifted
features, as shown in Fig.~\ref{g76_rel_pro}. We average the
absolute proper motions of both groups as the source's proper
motion.

\Gk: We detected only 3 features persisting over at least 3 epochs,
1 of which lasted 6 epochs and it was used for the parallax fit. The
amplitude of the internal proper motion is $\sim$10~\kms\ (see
Fig.~\ref{g79_rel_pro}). The absolute proper motion of the feature
employed for the parallax fit is taken to be the source's proper
motion.

\Gh: We detected only 2 features persisting over at least 3 epochs,
1 of which lasted 6 epochs and it is used for the parallax fit. The
other feature shows low relative velocity ($\sim$10~\kms), as shown
in Fig.~\ref{g90_rel_pro}. The absolute proper motion of the feature
employed for the parallax fit is taken to be the source's proper
motion.

\Gi: Although 27 features persisting over at least 3 epochs were
detected, only 2 of them persisted for 6 epochs and could be used
for the parallax fit. The internal motions of these features are
very complex (Fig.~\ref{g92_rel_pro}). We average all the
measured absolute proper motions as the source's proper motion.

\Gl: We detected only two features persisting for at least 3 epochs,
one of which lasted for 5 epochs and was used for the parallax fit.
The other feature had low relative velocity ($\sim$13~\kms), as
shown in Fig.~\ref{g105_rel_pro}. We only used the feature employed
for the parallax fit to derived the source's proper motion.

\begin{longtable}{rrrccc}
\caption{\label{kstars} Detailed Results of Parallax and Proper Motion
measurements.}\\
\hline \hline
Background & \Vlsr & Detected & Parallax & $\mu_x$ & $\mu_y$ \\
 Source & (\kms) & epochs & (mas) & (mas~yr$^{-1}$) & (mas~yr$^{-1}$)\\
\hline
\endfirsthead
\caption{continued.}\\
\hline\hline
Background & \Vlsr & Detected & Parallax & $\mu_x$ & $\mu_y$ \\
 Source & (\kms) & epochs & (mas) & (mas~yr$^{-1}$) & (mas~yr$^{-1}$)\\
\hline
\endhead
\hline
\endfoot
\multicolumn{6}{c}{\bf ON1}\\\hline
J2003+3034 & $+$14.8 & 111111 & 0.397$\pm$0.045 &$-$3.53$\pm$0.07
& $-$5.41$\pm$0.04 \\
           & $+$10.6 & 111111 & 0.421$\pm$0.018 &$-$2.97$\pm$0.02
& $-$5.60$\pm$0.06 \\
           & $+$10.6 & 011111 & 0.431$\pm$0.036 &$-$3.16$\pm$0.05
& $-$5.61$\pm$0.07 \\
\hline
\multicolumn{3}{c}{Combined fit} &  0.425$\pm$0.021  &   &  \\
\multicolumn{3}{c}{Average} &       & $-$3.22$\pm$0.05  & $-$5.54$\pm$0.06
\\\\
\hline \multicolumn{6}{c}{\bf \Ge} \\ \hline
\multicolumn{6}{c}{\bf center group} \\
J2025+3343 & $+$12.5 & 111111 & 0.665$\pm$0.010 &$-$3.59$\pm$0.05
& $-$3.55$\pm$0.07 \\
           & $+$12.9 & 111111 & 0.602$\pm$0.015 &$-$3.65$\pm$0.05
& $-$3.76$\pm$0.07 \\
           & $+$13.4 & 111111 & 0.624$\pm$0.012 &$-$3.48$\pm$0.05
& $-$3.68$\pm$0.07 \\
           & $+$6.9  & 001110 &                 &$-$7.30$\pm$0.48
& $-$4.20$\pm$0.43 \\
\multicolumn{3}{c}{Average}  &                  &$-$4.51$\pm$0.16
& $-$3.80$\pm$0.16 \\
\multicolumn{6}{c}{\bf Southeast group} \\
           & $+$0.8  & 111110 &                &$-$3.57$\pm$0.09
& $-$5.58$\pm$0.14 \\
           & $-$1.1  & 111000 &                &$-$6.04$\pm$0.24
& $-$9.71$\pm$0.33 \\
           & $+$2.4  & 111110 &                &$-$3.89$\pm$0.07
& $-$6.45$\pm$0.31 \\
           & $+$4.3  & 011111 &                &$-$2.72$\pm$0.30
& $-$5.97$\pm$0.20 \\
           & $+$2.1  & 011100 &                &$-$5.51$\pm$0.05
& $-$6.27$\pm$0.21 \\
           & $+$1.6  & 001110 &                &$-$6.33$\pm$0.28
& $-$11.0$\pm$1.87 \\
           & $+$3.1  & 000111 &                &$-$3.99$\pm$0.45
& $-$5.74$\pm$0.28 \\
\multicolumn{3}{c}{Average}  &                 &$-$4.58$\pm$0.21
& $-$7.25$\pm$0.48 \\
\multicolumn{6}{c}{\bf Northwest group} \\
           & $+$5.7  & 111000 &                &$-$3.50$\pm$0.15
& $-$3.39$\pm$0.14 \\
           & $+$7.0  & 001110 &                &$-$1.05$\pm$0.16
& $-$3.77$\pm$0.08 \\
\multicolumn{3}{c}{Average}  &                 &$-$2.28$\pm$0.16
& $-$3.58$\pm$0.11 \\
\hline
\multicolumn{3}{c}{Combined fit} &  0.629$\pm$0.010  &   &    \\
\multicolumn{3}{c}{Average} &       & $-$3.79$\pm$0.18  & $-$4.88$\pm$0.25
\\\\
\hline \multicolumn{6}{c}{\bf \Gf} \\ \hline
J2015+3710& $-$11.3 & 111111 & 0.285$\pm$0.004   &$-$2.09$\pm$0.01
& $-$4.76$\pm$0.15 \\
          & $-$10.9 & 011111 &                   &$-$2.05$\pm$0.09
& $-$4.92$\pm$0.04 \\
          & $-$10.4 & 111111 & 0.299$\pm$0.012   &$-$3.44$\pm$0.04
& $-$4.43$\pm$0.08 \\
          & $-$9.6  & 111111 & 0.291$\pm$0.017   &$-$3.34$\pm$0.04
& $-$5.26$\pm$0.08 \\
          & $-$9.2  & 111111 & 0.307$\pm$0.017   &$-$2.81$\pm$0.04
& $-$4.05$\pm$0.09 \\
          & $-$8.8  & 111100 &                   &$-$3.84$\pm$0.18
& $-$4.77$\pm$0.09 \\
          & $-$7.9  & 111101 &                   &$-$2.86$\pm$0.05
& $-$4.04$\pm$0.10 \\
          & $-$7.1  & 111111 & 0.236$\pm$0.033   &$-$4.42$\pm$0.14
& $-$3.52$\pm$0.06 \\
          & $-$5.0  & 111111 & 0.268$\pm$0.010   &$-$2.74$\pm$0.03
& $-$4.09$\pm$0.05\\
          & $-$4.6  & 111100 &                   &$-$3.22$\pm$0.01
& $-$5.78$\pm$0.10 \\
\hline
\multicolumn{3}{c}{Combined fit} & 0.285$\pm$0.009  &     &     \\
\multicolumn{3}{c}{Average} &       & $-$3.08$\pm$0.06 & $-$4.56$\pm$0.08
\\\\
\hline \multicolumn{6}{c}{\bf \Gm} \\ \hline
J2015+3710& $-$8.4 & 1110 &                   &$-$5.10$\pm$0.02
& $-$4.80$\pm$0.05 \\
          & $-$5.9 & 0111 &                   &$-$2.23$\pm$0.04
& $-$3.94$\pm$0.07 \\
          & $-$3.8 & 1011 &                   &$-$2.94$\pm$0.08
& $-$4.68$\pm$0.08 \\
          & $-$2.5 & 0111 &                   &$-$1.89$\pm$0.03
& $-$5.48$\pm$0.06 \\
          & $+$0.0 & 1110 &                   &$-$2.87$\pm$0.08
& $-$4.78$\pm$0.07 \\
          & $+$0.4 & 1111 & 0.258$\pm$0.015   &$-$1.86$\pm$0.02
& $-$5.13$\pm$0.03 \\
          & $+$0.4 & 1011 &                   &$-$1.77$\pm$0.08
& $-$5.02$\pm$0.14 \\
          & $+$1.3 & 0111 &                   &$-$2.98$\pm$0.13
& $-$5.53$\pm$0.09 \\
          & $+$1.7 & 1110 &                   &$-$3.37$\pm$0.02
& $-$3.99$\pm$0.03 \\
          & $+$1.7 & 1110 &                   &$-$1.64$\pm$0.19
& $-$4.79$\pm$0.03 \\
          & $+$2.1 & 0111 &                   &$-$1.98$\pm$0.16
& $-$4.91$\pm$0.17 \\
          & $+$2.5 & 0111 &                   &$-$3.46$\pm$0.05
& $-$4.23$\pm$0.10 \\
          & $+$2.5 & 1110 &                   &$-$1.61$\pm$0.05
& $-$5.25$\pm$0.06 \\
          & $+$2.9 & 1111 & 0.301$\pm$0.056   &$-$3.08$\pm$0.06
& $-$4.94$\pm$0.11 \\
          & $+$3.4 & 1111 & 0.310$\pm$0.044   &$-$2.28$\pm$0.05
& $-$4.43$\pm$0.06 \\
          & $+$3.4 & 0111 &                   &$-$3.23$\pm$0.02
& $-$4.70$\pm$0.01 \\
          & $+$3.8 & 1111 & 0.270$\pm$0.026   &$-$3.49$\pm$0.03
& $-$2.87$\pm$0.04 \\
          & $+$4.2 & 1111 & 0.289$\pm$0.019   &$-$2.91$\pm$0.02
& $-$4.88$\pm$0.02 \\
          & $+$4.6 & 1110 &                   &$-$3.11$\pm$0.03
& $-$4.99$\pm$0.15 \\
          & $+$9.3 & 1110 &                   &$-$3.92$\pm$0.19
& $-$5.09$\pm$0.04 \\
\hline
\multicolumn{3}{c}{Combined fit} & 0.281$\pm$0.015  &     &     \\
\multicolumn{3}{c}{Average} &       & $-$2.79$\pm$0.07  & $-$4.72$\pm$0.07
\\\\
\hline \multicolumn{6}{c}{\bf \Gg} 
\\ \hline
\multicolumn{6}{c}{\bf Northeast redshift group} \\
J2015+3710 & $+$6.9 & 111100 & 0.770$\pm$0.053 &$+$6.64$\pm$0.41
& $-$0.11$\pm$0.12 \\
\multicolumn{6}{c}{\bf Southwest blueshift group} \\
           & $-$13.3& 011100 &                 &$-$11.21$\pm$0.88
& $-$6.34$\pm$0.20 \\ 
           & $-$13.3& 011100 &                 &$-$16.98$\pm$4.70
& $-$8.78$\pm$1.23 \\ 
\multicolumn{3}{c}{Average} &              & $-$14.10$\pm$2.79
& $-$7.56$\pm$0.72 \\
\hline
\multicolumn{3}{c}{Combined fit} & 0.770$\pm$0.053  &     &     \\
\multicolumn{3}{c}{Average} &       & $-$3.73$\pm$1.60  & $-$3.84$\pm$0.42
\\\\
\hline \multicolumn{6}{c}{\bf \Gk} \\ \hline
J2007+4029 & $-$4.6 & 1111110 & 0.620$\pm$0.027 &$-$3.23$\pm$0.07
& $-$5.19$\pm$0.15 \\\\
\hline \multicolumn{6}{c}{\bf \Gh} \\ \hline
J2056+4940 & $-$6.2 & 111111& 1.487$\pm$0.064 &$-$0.66$\pm$0.18
& $-$0.93$\pm$1.11 \\
J2059+4851 & $-$6.2 & 111111& 1.481$\pm$0.081 &$-$0.68$\pm$0.18
& $-$0.76$\pm$1.11 \\
J2114+4953 & $-$6.2 & 111111& 1.481$\pm$0.079 &$-$0.68$\pm$0.18
& $-$1.02$\pm$1.11 \\
\hline
\multicolumn{3}{c}{Combined fit} & 1.483$\pm$0.038 & $-$0.67$\pm$0.10
& $-$0.90$\pm$0.59    \\\\
\hline \multicolumn{6}{c}{\bf \Gi} \\ \hline
J2117+5431 & $+$0.1  & 111000 &                 &$-$2.12$\pm$0.01
& $-$4.56$\pm$0.28 \\
           & $-$0.5  & 111000 &                 &$+$2.46$\pm$2.26
& $-$1.76$\pm$0.84 \\
           & $-$1.1  & 011110 &                 &$-$0.99$\pm$0.31
& $-$2.67$\pm$0.27 \\
           & $-$1.8  & 111110 &                 &$-$2.44$\pm$0.11
& $-$4.28$\pm$0.11 \\
           & $-$2.5  & 000111 &                 &$-$1.95$\pm$0.22
& $-$3.29$\pm$0.31 \\
           & $-$3.5  & 001110 &                 &$-$2.15$\pm$0.35
& $-$2.76$\pm$0.52 \\
           & $-$3.7  & 111111 & 0.635$\pm$0.010 &$-$1.70$\pm$0.03
& $-$3.15$\pm$0.08 \\
           & $-$4.9  & 011111 &                 &$-$1.83$\pm$0.16
& $-$4.68$\pm$0.22 \\
           & $-$6.2  & 011110 &                 &$-$1.74$\pm$0.11
& $-$3.65$\pm$0.26 \\
           & $-$6.5  & 111100 &                 &$-$2.01$\pm$0.07
& $-$4.57$\pm$0.03 \\
           & $-$6.8  & 001110 &                 &$-$2.70$\pm$0.05
& $-$4.47$\pm$0.44 \\
           & $-$7.4  & 111100 &                 &$-$2.39$\pm$0.14
& $-$4.41$\pm$0.14 \\
           & $-$8.3  & 111000 &                 &$-$1.18$\pm$0.28
& $-$3.21$\pm$0.38 \\
           & $-$10.0 & 111000 &                 &$-$1.47$\pm$0.01
& $+$0.41$\pm$0.01 \\
           & $-$10.7 & 111000 &                 &$-$0.17$\pm$0.12
& $-$2.69$\pm$0.32 \\
           & $-$12.8 & 111000 &                 &$+$1.24$\pm$0.36
& $-$2.25$\pm$1.13 \\
           & $-$13.7 & 001110 &                 &$+$0.21$\pm$0.05
& $-$5.12$\pm$0.47 \\
           & $-$14.3 & 011110 &                 &$+$0.95$\pm$0.07
& $-$1.07$\pm$0.23 \\
           & $-$14.3 & 111111 & 0.593$\pm$0.023 &$-$0.03$\pm$0.06
& $-$1.81$\pm$0.25 \\
           & $-$14.5 & 111000 &                 &$+$1.47$\pm$0.01
& $-$1.21$\pm$0.12 \\
           & $-$15.2 & 111000 &                 &$+$0.90$\pm$0.51
& $-$0.73$\pm$0.09 \\
           & $-$16.2 & 111110 &                 &$+$0.82$\pm$0.30
& $-$1.45$\pm$0.64 \\
           & $-$26.6 & 011110 &                 &$-$1.59$\pm$0.27
& $-$1.85$\pm$0.45 \\
           & $-$29.5 & 111000 &                 &$+$1.49$\pm$0.79
& $+$0.04$\pm$0.42 \\
           & $-$33.7 & 001011 &                 &$-$1.55$\pm$0.12
& $+$2.03$\pm$0.15 \\
           & $-$36.5 & 111000 &                 &$+$0.41$\pm$0.21
& $+$0.68$\pm$0.71 \\
           & $-$37.4 & 001110 &                 &$-$0.55$\pm$0.02
& $+$1.82$\pm$0.06 \\

\hline
\multicolumn{3}{c}{Combined fit} & 0.613$\pm$0.014  &     &     \\
\multicolumn{3}{c}{Average} &       & $-$0.69$\pm$0.26 & $-$2.25$\pm$0.33
\\\\
\hline \multicolumn{6}{c}{\bf \Gl} \\  \hline
J2203+6750 & $-$12.1 & 0011111 & 1.129$\pm$0.063 &$-$0.21$\pm$0.24
& $-$5.49$\pm$0.13 \\
\hline
\label{table:detail}
\end{longtable}

\begin{figure}
\includegraphics[angle=-90,scale=0.6]{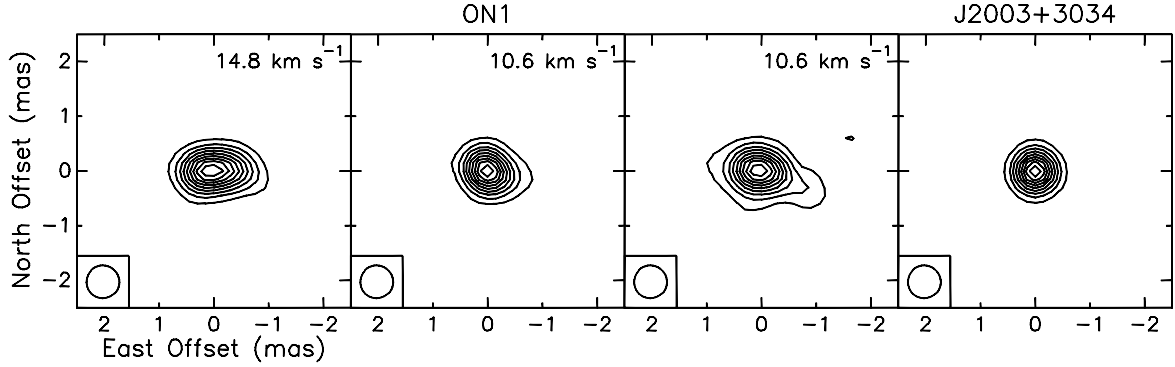}
\caption{\scriptsize Maser channel-maps of ON~1 and the
phase-reference source J2003$+$3034 for the second epoch (2009 May
7) (the spot at \Vlsr = 10.6~\kms\ was not detected at the first
epoch). The channel velocities and restoring beams are given in the
upper right and lower left corner of each panel. Contour levels are
integer multiples of 10\% of the peak brightness of 34.1, 5.9 and
3.6~Jy~beam$^{-1}$ for ON 1 (from left to right) and
0.1~Jy~beam$^{-1}$ for J2003$+$3034. The two spots at \Vlsr =
10.6~\kms\ have a separation of $\sim$35 mas. They appear to be
compact and excellent astrometric targets. \label{on1_masers}}
\end{figure}

\begin{figure}
\includegraphics[angle=-90,scale=0.64]{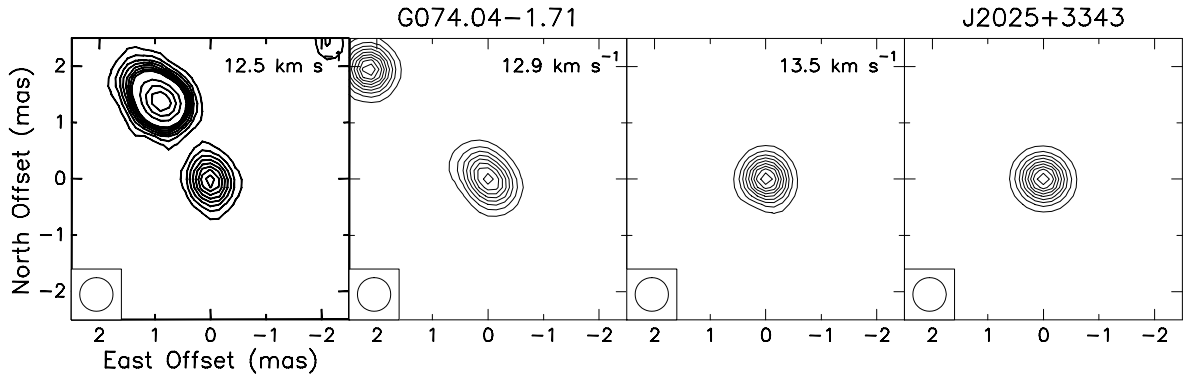}
\caption{\scriptsize Maser channel-maps of \Ge\ and the
phase-reference source J2025$+$3343 for the first epoch (2010 April
24). Contour levels are integer multiples of 10\% of the peak
brightness of 0.3, 0.5 and 0.6~Jy~beam$^{-1}$ for \Ge\ (from left to
right) and 3.9~Jy~beam$^{-1}$ for J2025$+$3343. The maser spots at
the center of the images are used for the parallax fit.
\label{g74_masers}}
\end{figure}

\begin{figure}
\includegraphics[angle=-90,scale=0.62]{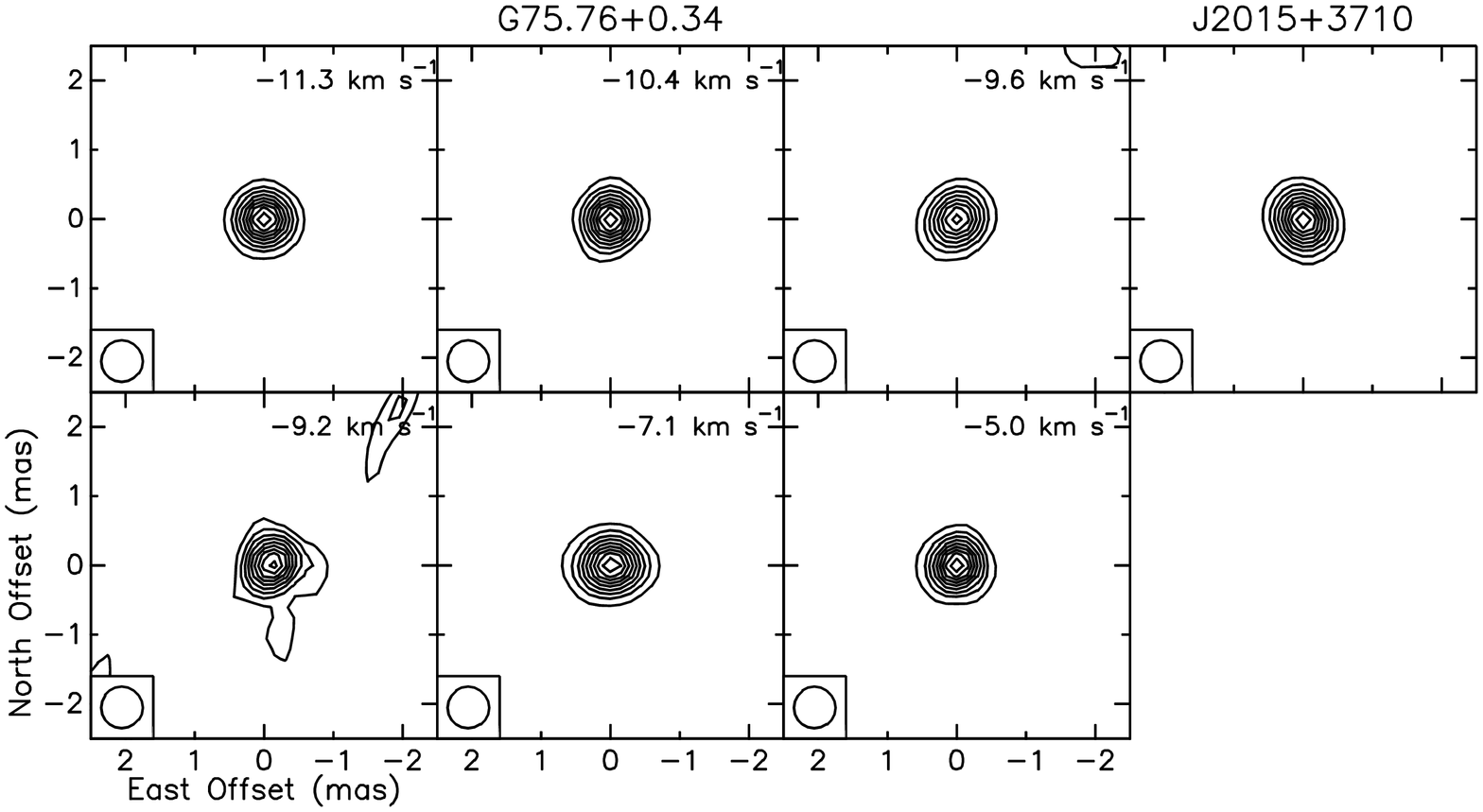}
\caption{\scriptsize Maser channel-maps of \Gf\ and the
phase-reference source J2015$+$3710 for the first epoch (2010 April
24). Contour levels are integer multiples of 10\% of the peak
brightness of 3.3, 2.4, 5.3, 0.5, 1.4 and 0.7~Jy~beam$^{-1}$ for
\Gf\ (from upper left to lower right) and 3.9~Jy~beam$^{-1}$ for
J2015$+$3710. \label{g75_masers}}
\end{figure}

\begin{figure}
\includegraphics[angle=-90,scale=0.45]{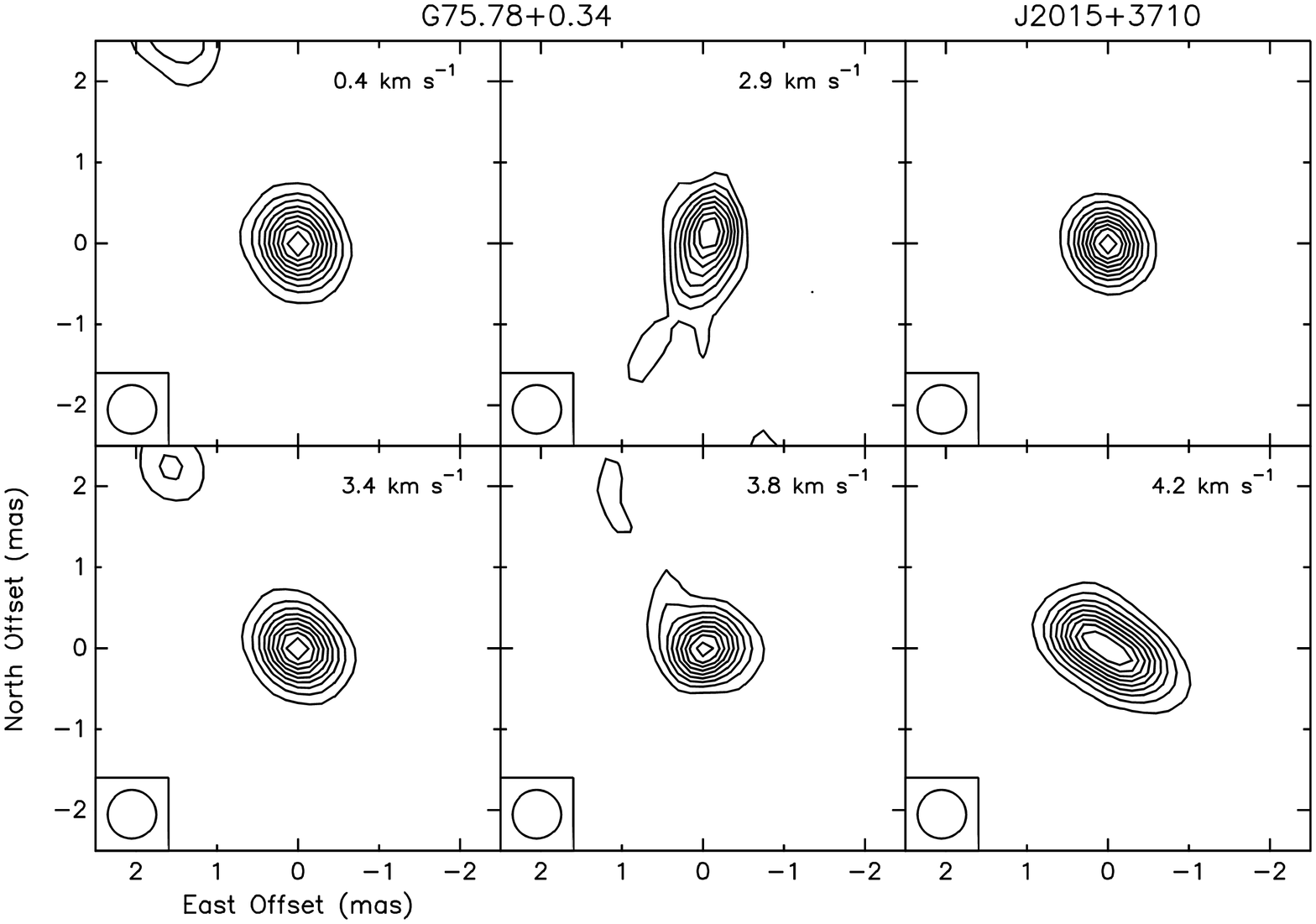}
\caption{\scriptsize Maser channel-maps of \Gm\ and the
phase-reference source J2015$+$3710 for the first epoch (2009 May
7). Contour levels are integer multiples of 10\% of the peak
brightness of 26.7, 2.4, 58.8, 6.0 and 10.8~Jy~beam$^{-1}$ for \Gm\
(from upper left to lower right) and 2.3~Jy~beam$^{-1}$ for
J2015$+$3710. \label{g75_masers}}
\end{figure}

\begin{figure}
\includegraphics[angle=-90,scale=0.37]{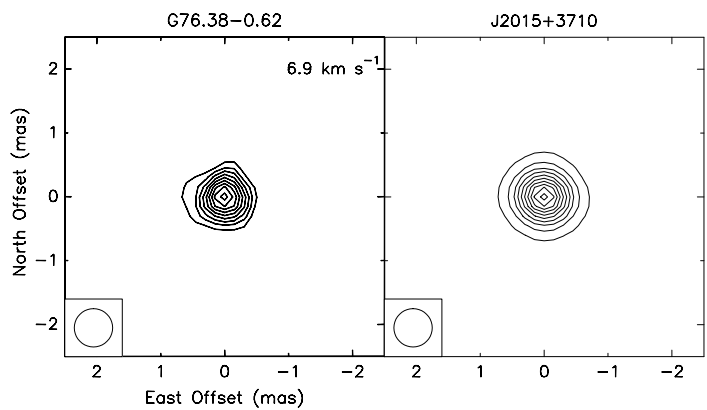}
\caption{\scriptsize Maser channel-map at \Vlsr = 6.9~\kms\ of \Gg\ and
the phase-reference source J2015$+$3710 for the first epoch (2010
April 24). Contour levels are integer multiples of 10\% of the
peak brightness of 0.2~Jy~beam$^{-1}$ for \Gg\ and
1.6~Jy~beam$^{-1}$ for J2015$+$3710. \label{g76_masers}}
\end{figure}

\begin{figure}
\includegraphics[angle=-90,scale=0.42]{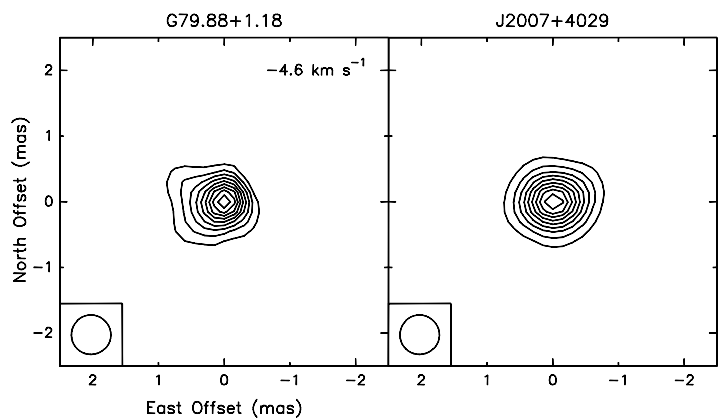}
\caption{\scriptsize Maser channel-map at \Vlsr = -4.6~\kms\ of \Gk\
and the phase-reference source J2007$+$4029 for the first epoch
(2011 May 24). Contour levels are integer multiples of 10\% of
the peak brightness of 8.0 for \Gk,\ and 1.6~Jy~beam$^{-1}$
J2007$+$4029, respectively. \label{g79_masers}}
\end{figure}

\begin{figure}
\includegraphics[angle=-90,scale=0.65]{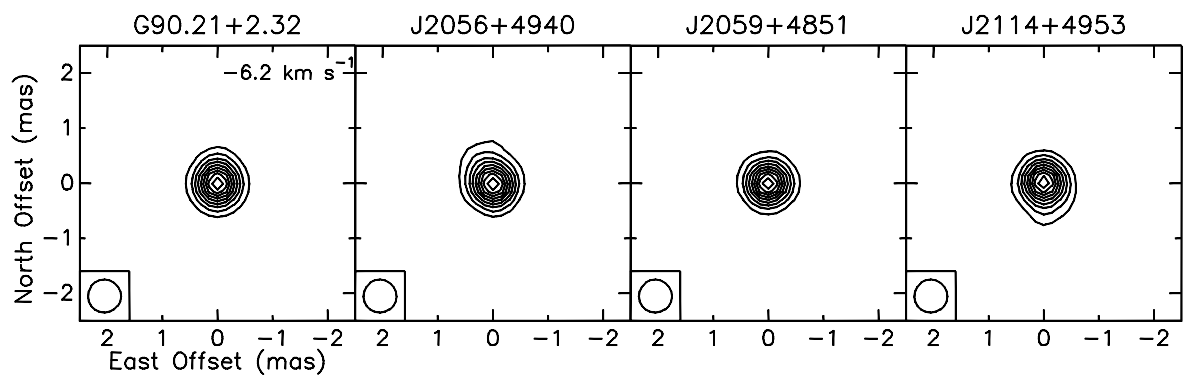}
\caption{\scriptsize Image of the maser channel at \Vlsr = -6.2~\kms\ of \Gh\
and the background sources J2056$+$4940, J2059$+$4851 and
J2114$+$4953 for the first epoch (2010 May 16). Contour levels are
at integer multiples of 10\% of the peak brightness of
17.0~Jy~beam$^{-1}$ for \Gh\ and 0.03, 0.05 and 0.06~Jy~beam$^{-1}$
for J2056$+$4940, J2059$+$4851 and J2114$+$4953, respectively.
\label{g90_masers}}
\end{figure}

\begin{figure}
\includegraphics[angle=-90,scale=0.5]{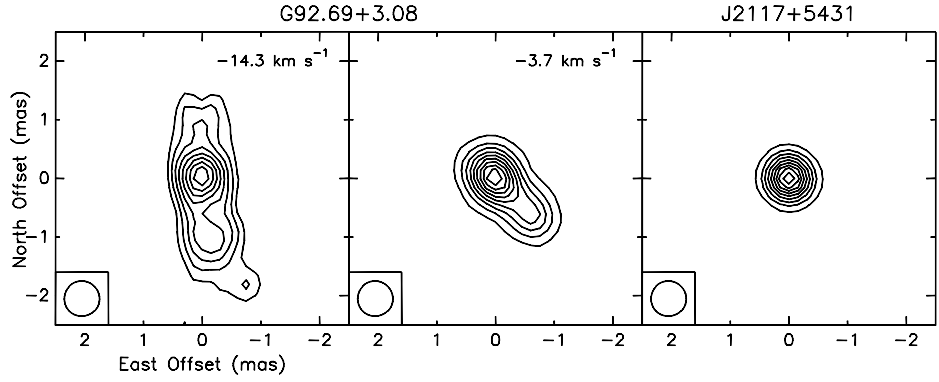}
\caption{\scriptsize Maser channel-maps of \Gi\ and the phase-reference
source J2117$+$5431 for the first epoch (2010 May 16). 
Contour levels are integer multiples of 10\% of the peak
brightness of 26.1 and 16.4~Jy~beam$^{-1}$ (from left to right) for
\Gi\ and 0.2~Jy~beam$^{-1}$ for J2117$+$5431. \label{g92_masers}}
\end{figure}

\begin{figure}
\includegraphics[angle=-90,scale=0.4]{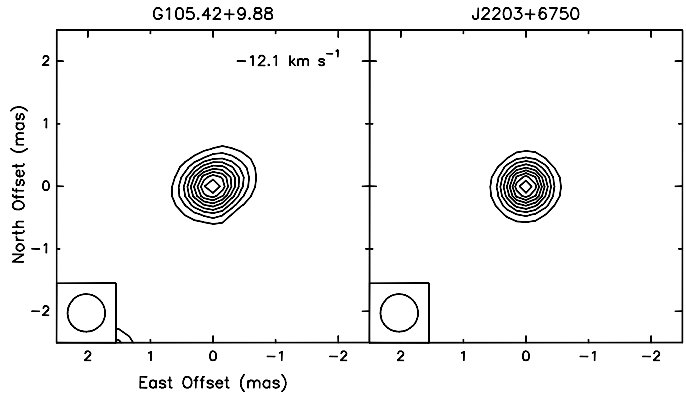}
\caption{\scriptsize Maser channel-map at \Vlsr = -12.1~\kms\ of \Gl\
and the phase-reference source J2203$+$6750 for the third epoch
(2011 October 30). Contour levels are integer multiples of 10\%
of the peak brightness of 2.2~Jy~beam$^{-1}$ for \Gl\ and
0.2~Jy~beam$^{-1}$ for J2203$+$6750, respectively.
\label{g105_masers}}
\end{figure}

\clearpage

\begin{figure}
\includegraphics[scale=0.4]{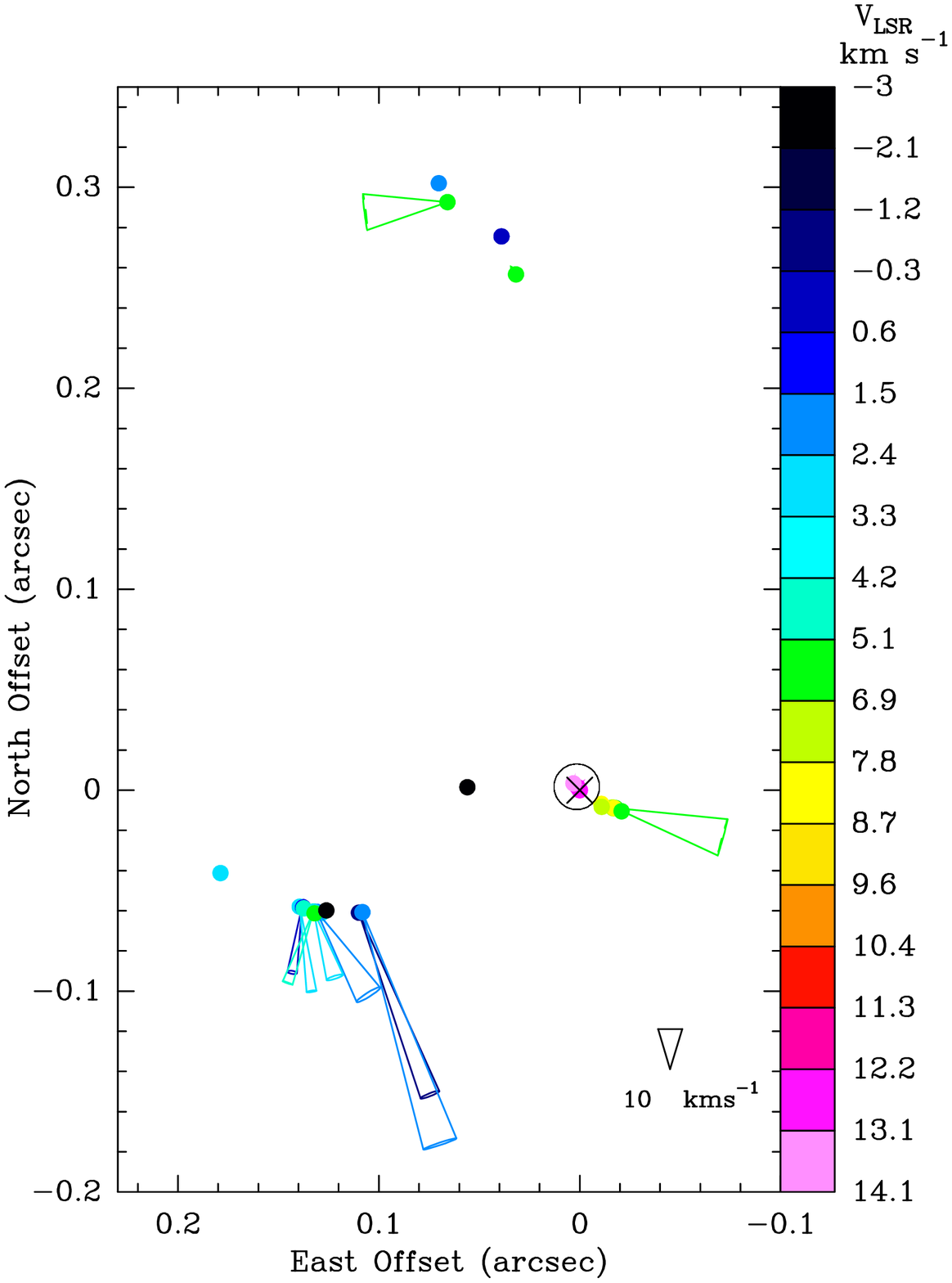}
\caption{\scriptsize Distributions of \hho\ masers in \Ge.
Maser LSR velocities are indicated by
the color scale on the righthand side of the plot. The
cones indicate the 3-D velocities relative to the persistent spot
(marked with a cross) at \Vlsr = 13.4 \kms: R.A.(J2000) = $20^{\rm
h}25^{\rm m}07.1053^{\rm s}$ and Dec.(J2000) =
$34\degr49\arcmin57.593\arcsec$ on 2010 April 24. The cone opening
angle gives the 1$\sigma$ uncertainty on the proper motion
direction. The length of the cone is proportional to the velocity,
with a 10 \kms\ scale indicated in the lower right corner.
Points without associated cones were detected in fewer than three
epochs.
The maser features used for the determination of the parallax are
marked with black circles. \label{g74_rel_pro}}
\end{figure}

\begin{figure}
\includegraphics[angle=-90,scale=0.4]{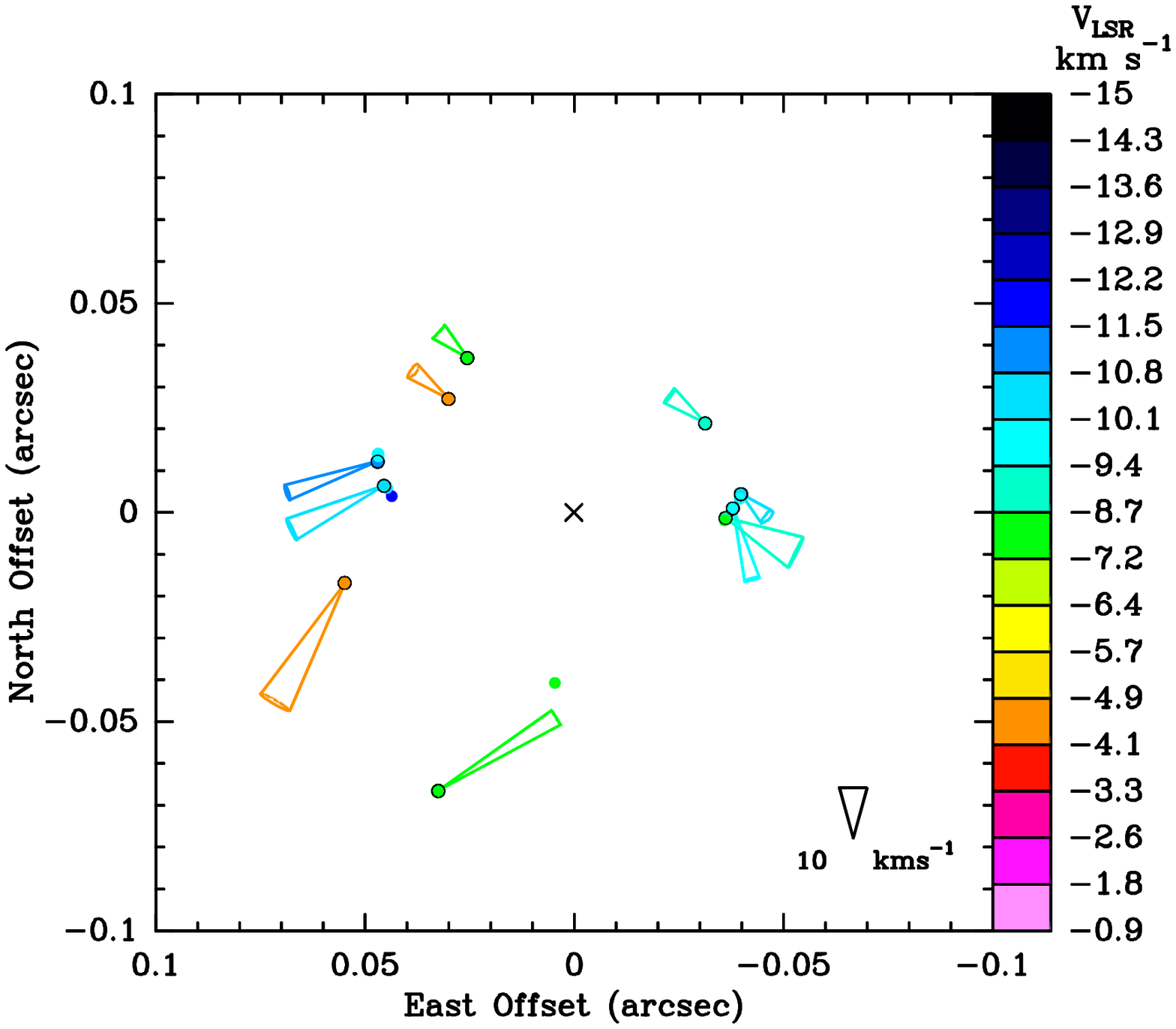}
\caption{\scriptsize Distributions of \hho\ masers in \Gf. Symbols
and cones have the same meaning as in Fig.~\ref{g74_rel_pro}, except
that the proper motions are relative to the the geometric center (x) of the
features persisting
over the 6 observing epochs: R.A.(J2000) = $20^{\rm h}21^{\rm m}41.0894^{\rm s}$
and Dec.(J2000) = $37\degr25\arcmin29.274\arcsec$
on 2010 April 24. The maser features used for the determination of
the parallax and source's proper motion are marked with black
circles. \label{g75_rel_pro}}
\end{figure}

\begin{figure}
\includegraphics[scale=0.4]{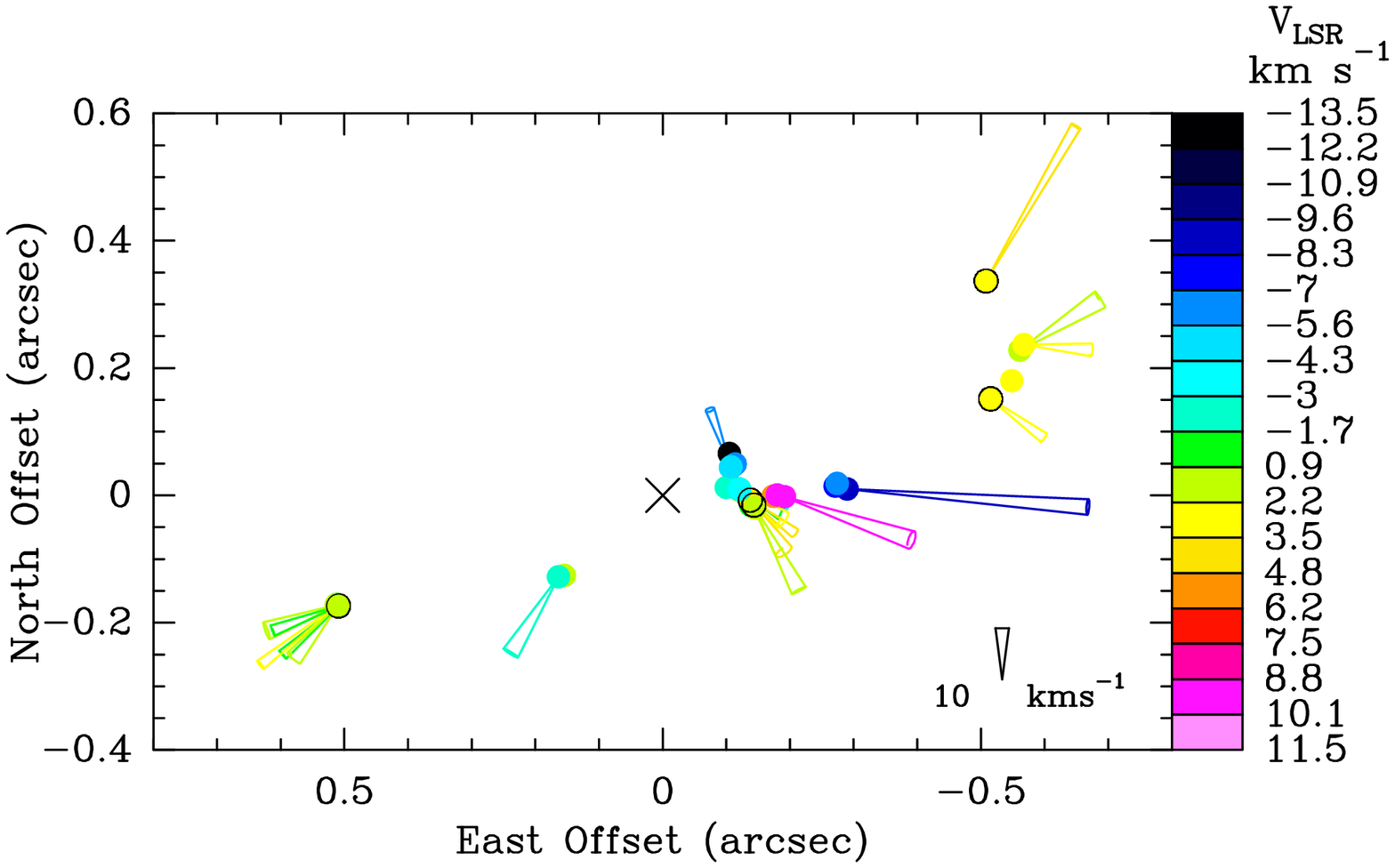}
\caption{\scriptsize Distributions of \hho\ masers in \Gm. Symbols
and cones have the same meaning as in Fig.~\ref{g74_rel_pro}, except
that the proper motions are relative to the geometric center (x) of the
features persisting
over the 4 observing epochs: R.A.(J2000) = $20^{\rm h}21^{\rm
m}44.0218^{\rm s}$ and Dec.(J2000) = $37\degr26\arcmin37.462\arcsec$
on 2009 May 7. The maser features used for the determination of
the parallax are marked with black circles. \label{g75.78_rel_pro}}
\end{figure}

\begin{figure}
\includegraphics[scale=0.4]{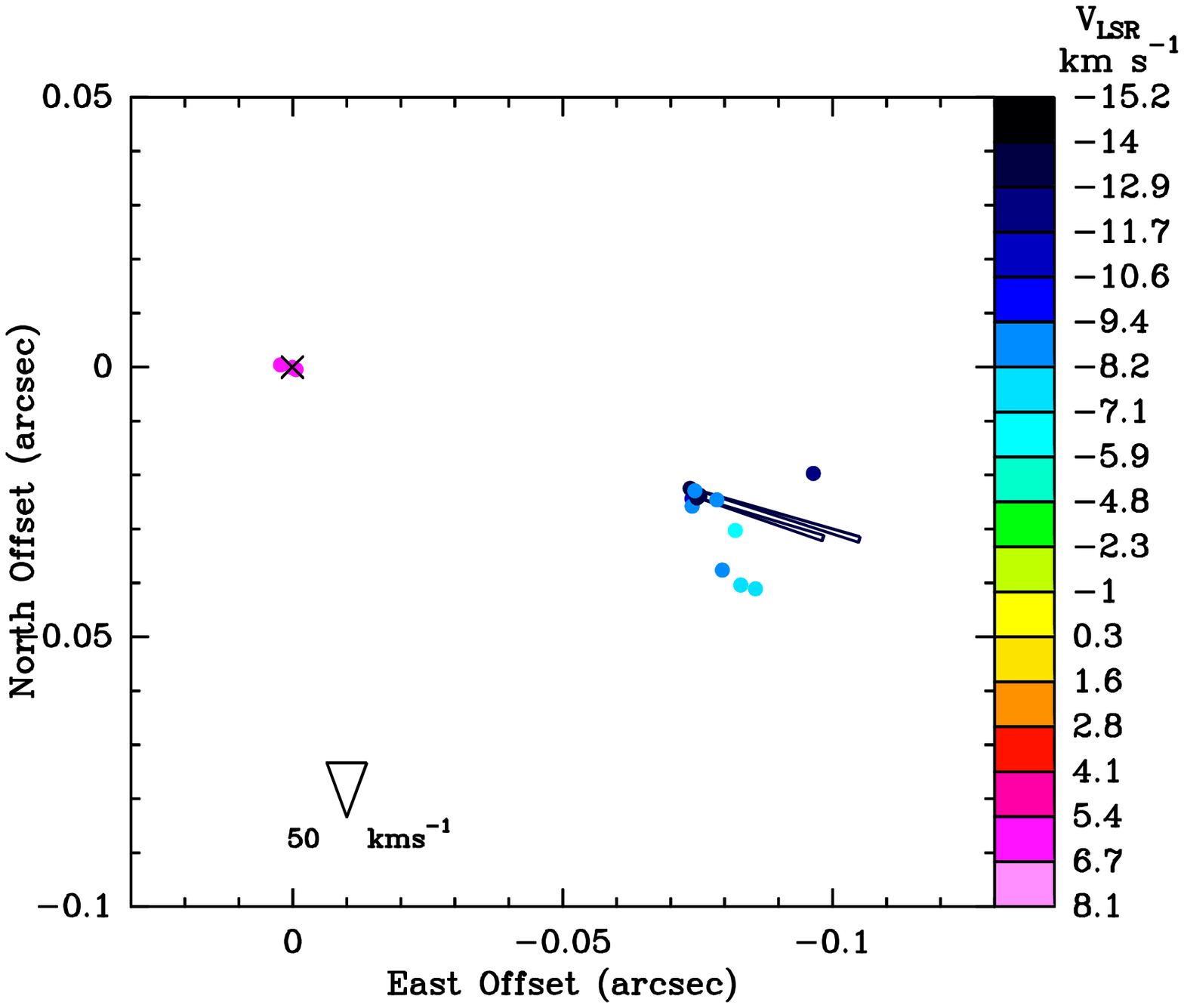}
\caption{\scriptsize Distributions of \hho\ masers in \Gg. Symbols
and cones have the same meaning as in Fig.~\ref{g74_rel_pro}. The
proper motions are relative to the persistent spot used for the
parallax fit (marked with a cross) at \Vlsr = 6.9 \kms: R.A.(J2000)
= $20^{\rm h}27^{\rm m}25.4816^{\rm s}$ and Dec.(J2000) =
$37\degr22\arcmin48.482\arcsec$ on 2010 April
24.\label{g76_rel_pro}}
\end{figure}

\begin{figure}
\includegraphics[scale=0.4]{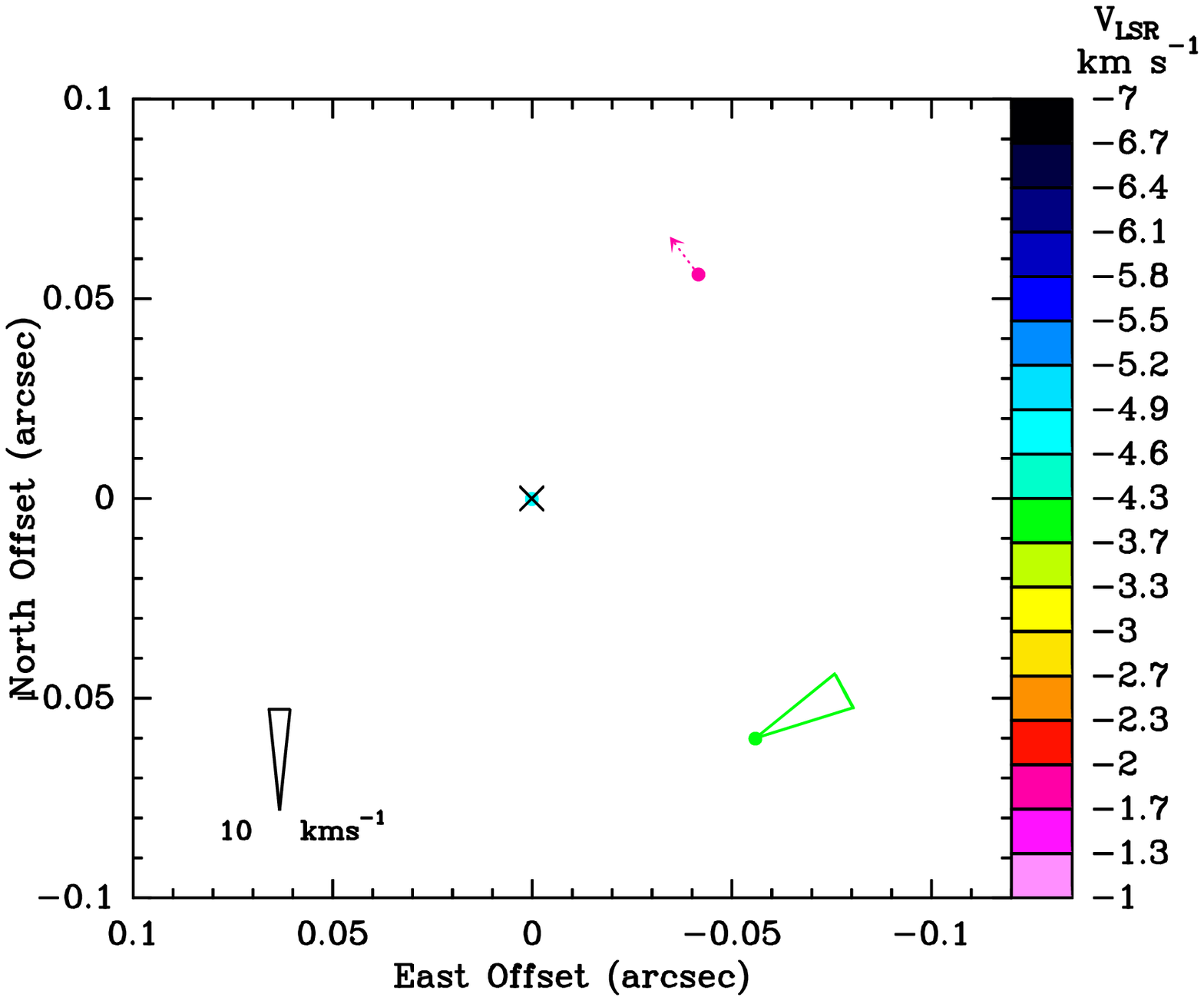}
\caption{\scriptsize Distributions of \hho\ masers in \Gk. Symbols
and cones have the same meaning as in Fig.~\ref{g74_rel_pro}. The
proper motions are relative to the persistent spot used for the
parallax fit (marked with a cross) at \Vlsr = -4.6 \kms: R.A.(J2000)
= $20^{\rm h}30^{\rm m}29.1464^{\rm s}$ and Dec.(J2000) =
$41\degr15\arcmin53.590\arcsec$ on 2011 May 24.\label{g79_rel_pro}}
\end{figure}

\begin{figure}
\includegraphics[scale=0.4]{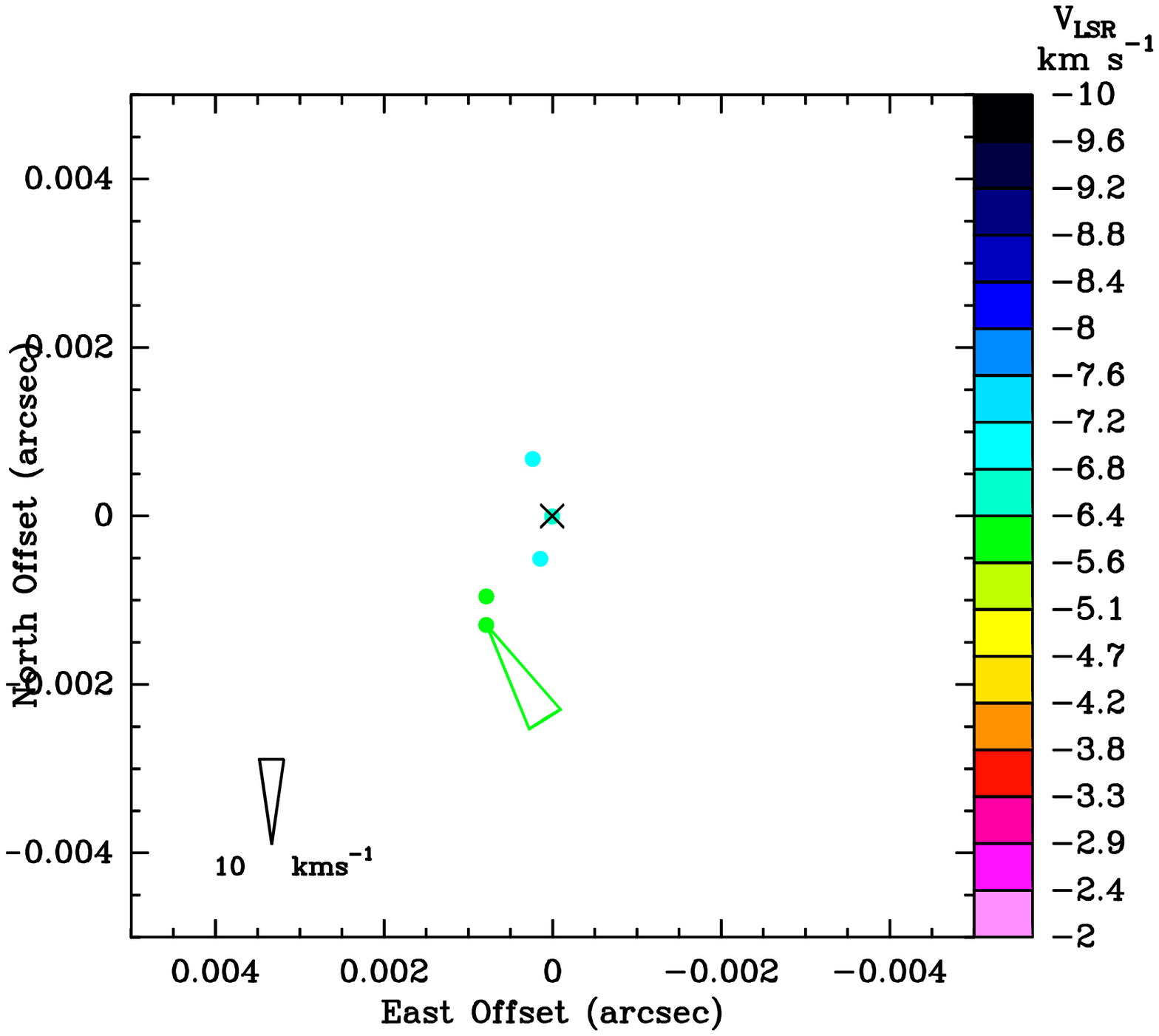}
\caption{\scriptsize Distributions of \hho\ masers in \Gh. Symbols
and cones have the same meaning as in Fig.~\ref{g74_rel_pro}. The
proper motions are relative to the persistent spot used for the
parallax fit (marked with a cross) at \Vlsr = -6.2 \kms: R.A.(J2000)
= $21^{\rm h}02^{\rm m}22.7007^{\rm s}$ and Dec.(J2000) =
$50\degr03\arcmin08.309\arcsec$ on 2010 May 16.\label{g90_rel_pro}}
\end{figure}

\begin{figure}
\includegraphics[scale=0.4]{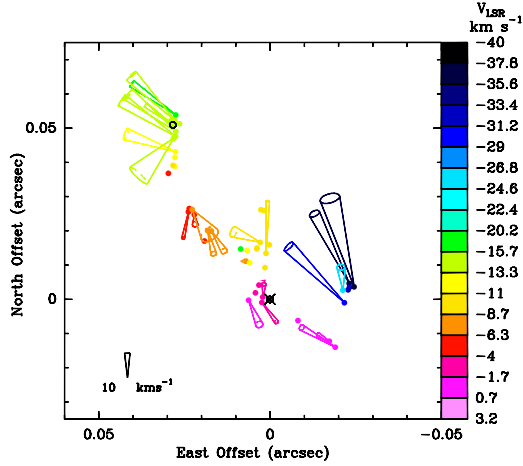}
\caption{\scriptsize Distributions of \hho\ masers in \Gi.
Symbols and cones have the same meaning as in
Fig.~\ref{g74_rel_pro}. The proper motions are relative to the
persistent spot (marked with a cross) at \Vlsr = -3.7 \kms:
R.A.(J2000) = $21^{\rm h}09^{\rm m}21.7232^{\rm s}$ and Dec.(J2000)
= $52\degr22\arcmin37.083\arcsec$ on 2010 May 16. The maser features
used for the determination of the parallax are marked with black
circles. \label{g92_rel_pro}}
\end{figure}

\begin{figure}
\includegraphics[scale=0.4]{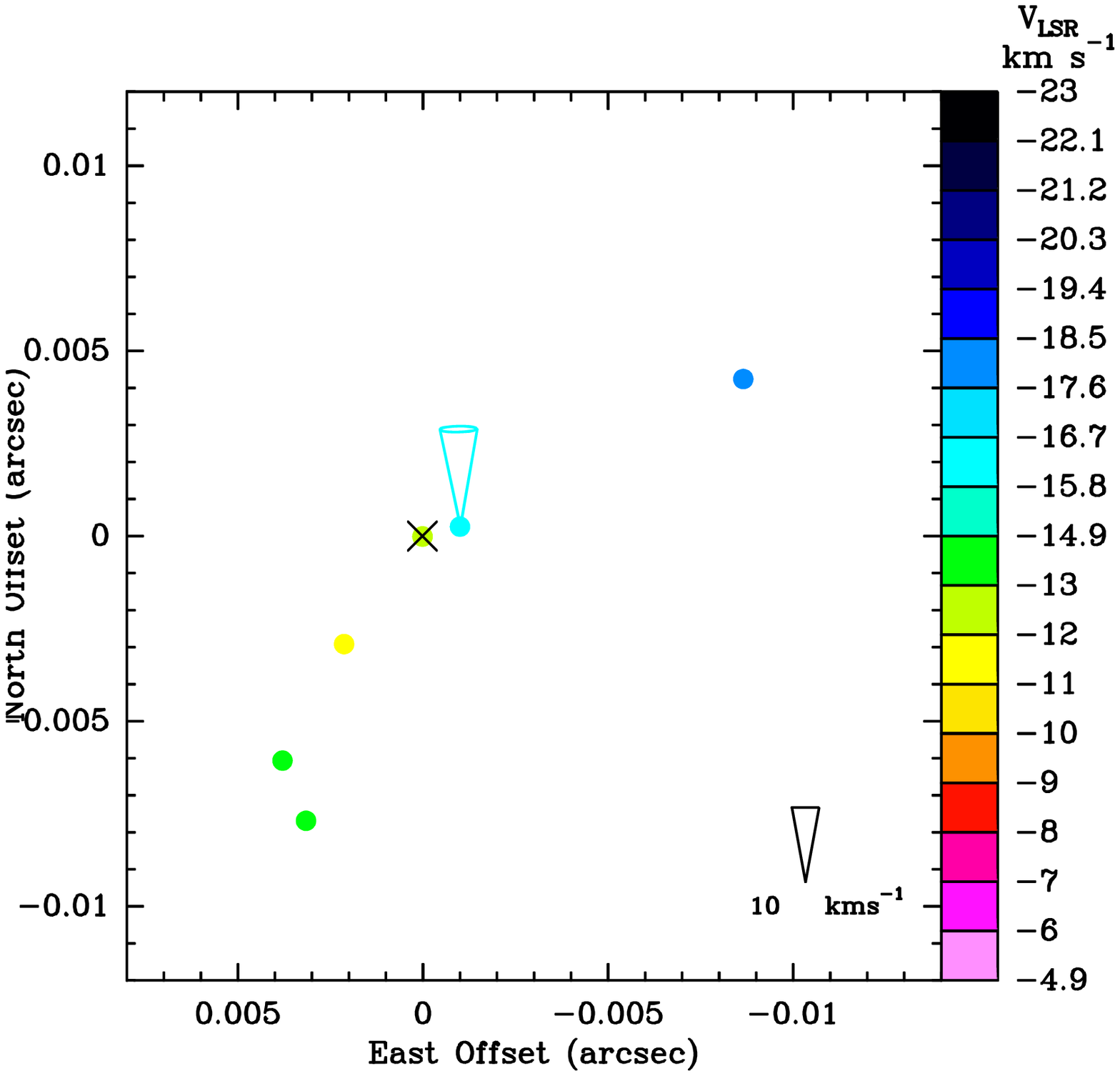}
\caption{\scriptsize Distributions of \hho\ masers in \Gl. Symbols
and cones have the same meaning as in Fig.~\ref{g74_rel_pro}. The
proper motions are relative to the persistent spot used for the
parallax fit (marked with a cross) at \Vlsr = -12.1 \kms:
R.A.(J2000) = $21^{\rm h}43^{\rm m}06.4628^{\rm s}$ and Dec.(J2000)
= $66\degr06\arcmin55.183\arcsec$ on 2011 October 30.
\label{g105_rel_pro}}
\end{figure}


\begin{thebibliography}{}

\bibitem[Ando et al.(2011)]{Ando:11} Ando, K., Nagayama, T., Omodaka,
T., et al. 2011, \pasj, 63, 45
\bibitem[Avedisova(1985)]{Avedisova:85} Avedisova, V. S. 1985, Soviet
Astronomy Letters, 11, 185
\bibitem[Bachiller et al.(1990)]{Bachiller:90} Bachiller, R., Martin-Pintado,
J., Tafalla, M., Cernicharo, J., \& Lazareff, B. 1990, \aap, 231, 174
\bibitem[Bok(1959)]{Bok:59} Bok, B. J. 1959, The Observatory, 79, 58
\bibitem[Bok et al.(1970)]{Bok:70} Bok, B. J., Hine, A. A., \& Miller,
E. W. 1970, in IAU Symp. 38, The Spiral Structure of our Galaxy,
ed. W. Becker \& G. Contopoulos, p.246
\bibitem[Bronfman et al.(1996)]{Bro:96} Bronfman, L., Nyman, L.-A.,
\& May, J. 1996, \aaps, 115, 81
\bibitem[Bronfman et al.(2000)]{Bro:00} Bronfman, L., May, J., \&
Luna, A. 2000, in ASP Conf. Ser. 217, Imaging at Radio through Submillimeter
Wavelengths, ed. J. G. Mangum \& S. J. E. Radford, p.66
\bibitem[Brunthaler et al.(2011)]{Brunthale11} Brunthaler, A., Reid,
M. J, Menten, K. M., et al. 2011, \textit{AN}, 332, 461
\bibitem[Carral et al.(1997)]{Carral:97} Carral, P., Kurtz, S. E.,
Rodr\'{\i}guez, L. F., De Pree, C., \& Hofner, P.
1997, \apj, 486, L103
\bibitem[Choi et al.(1999)]{Choi:99} Choi, M., Panis, J.-F., \& Evans,
II, N.~J. 1999, \apjs, 122, 519
\bibitem[Choi et al.(2008)]{Choi:08} Choi, Y. K., Hirota, T., Honma,
M., et al. 2008, \pasj, 60, 1007
\bibitem[Churchwell et al.(2009)]{Churchwell:09} Churchwell, E.,
Babler, B. L., Meade, M. R., et al. 2009, \pasp, 121, 213
\bibitem[Clark(1986)]{Clark:86} Clark, F. O. 1986, \aap, 164, L19
\bibitem[Codella et al.(2010)]{Codella:10} Codella, C., Cesaroni,
R., L\'{o}pez-Sepulcre, A., et al. 2010, \aap, 510, A86
\bibitem[Cordes \& Lazio(2002)]{Cordes:02} Cordes, J. M, \& Lazio,
T. J. W. 2002, arXiv:astro-ph/0207156
\bibitem[Dame et al.(2001)]{Dame:01} Dame, T. M., Hartmann, D., \&
Thaddeus, P. 2001, \apj, 547, 792
\bibitem[Deller et al.(2007)]{Del:07} Deller, A. T., Tingay, S. J.,
Bailes, M., \& West, C. 2007, \pasp, 119, 318D
\bibitem[Di et al.(2008)]{Di:08} Di Francesco, J., Johnstone, D.,
Kirk, H., MacKenzie, T., \& Ledwosinska, E. 2008, \apjs, 175, 277
\bibitem[Dickel et al.(1978)]{Dic:78} Dickel, J. R., Dickel, H. R.,
\& Wilson, W. J. 1978, \apj, 223, 840
\bibitem[Dehnen \& Binney(1998)]{Dehnen:98} Dehnen, W., \& Binney, J. J. 1998,
\mnras, 298, 387
\bibitem[Dzib et al.(2010)]{Dzib:10} Dzib, S., Loinard, L., Mioduszewski,
A. J., et al. 2010, \apj, 718, 610
\bibitem[Eiroa et al.(1979)]{Eiroa:79} Eiroa, C., Elsaesser, H.,
\& Lahulla, J. F. 1979, \aap, 74, 89
\bibitem[Elmegreen(1980)]{Elm:80} Elmegreen, D. M. 1980, \apj, 242, 528
\bibitem[Fomalont et al.(2003)]{Fomalont:03} Fomalont, E.~B., Petrov,
L., MacMillan, D.~S., Gordon, D., \& Ma, C. 2003, \aj, 126, 2562
\bibitem[Garay et al.(1993)]{Garay:93} Garay, G., Rodriguez, L. F.,
Moran, J. M., \& Churchwell, E. 1993, \apj, 418, 368
\bibitem[Georgelin \& Georgelin(1976)]{Georgelin:76} Georgelin,
Y. M., \& Georgelin, Y. P. 1976, \aap, 49, 57,
\bibitem[Harju et al.(1998)]{Harju:98} Harju, J., Lehtinen, K., Booth,
R. S., \& Zinchenko, I. 1998, \aaps, 132, 211
\bibitem[Hirota et al.(2011)]{Hirota:11} Hirota, T., Honma, M., Imai,
H., et al. 2011, \pasj, 63, 1
\bibitem[Hirota et al.(2008a)]{Hirota:08a} Hirota, T., Ando, K., Bushimata,
T., et al. 2008a, \pasj, 60, 961
\bibitem[Hirota et al.(2008b)]{Hirota:08b} Hirota, T., Bushimata,
T., Choi, Y. K., et al. 2008b, \pasj, 60, 37
\bibitem[Hirota et al.(2007)]{Hirota:07} Hirota, T., Bushimata, T.,
Choi, Y. K., et al. 2007, \pasj, 59, 897
\bibitem[Hofner \& Churchwell(1996)]{Hofner:96} Hofner, P., \& Churchwell,
E. 1996, \aaps, 120, 283
\bibitem[Honma et al.(2012)]{Honma:12} Honma, M., Nagayama, T., Ando, K.,
et al. 2012, \pasj, 64, 136H
\bibitem[Hou et al.(2009)]{Hou:09} Hou, L. G., Han, J. L., \& Shi,
W. B. 2009, \aap, 499, 473
\bibitem[Imai et al.(2007)]{Imai:07} Imai, H., Nakashima, K., Bushimata,
T., et al. 2007, \pasj, 59, 1107
\bibitem[Immer et al.(2011)]{Immer:11} Immer, K., Brunthaler, A.,
Reid, M. J., et al. 2011, \apjs, 194, 25
\bibitem[Kemper et al.(2003)]{Kemper:03} Kemper, F., Stark, R., Justtanont,
K., et al. 2003, \aap, 407, 609
\bibitem[Kennicutt(1981)]{Kennicutt:81} Kennicutt, R. C., Jr. 1981,
\aj, 86, 1847
\bibitem[Kerr(1970)]{Kerr:70} Kerr, F. J. 1970, in IAU Symp. 38, The
Spiral Structure of our Galaxy, ed. W. Becker \& G. Contopoulos, p.95
\bibitem[Kerr \& Kerr(1970)]{KK:70} Kerr, F. J., \& Kerr, M. 1970,
\apj, 6, 175
\bibitem[Kim et al.(2008)]{Kim:08} Kim, M. K., Hirota, T., Honma,
M., et al. 2008, \pasj, 60, 991
\bibitem[Kumar et al.(2004)]{Kumar:04} Kumar, M. S. N., Tafalla, M.,
\& Bachiller, R. 2004, \aap, 426, 195
\bibitem[Kurtz et al.(1994)]{Kurtz:94} Kurtz, S., Churchwell, E.,
\& Wood, D. O. S. 1994, \apjs, 91, 659
\bibitem[Kurtz et al.(2004)]{Kurtz:04} Kurtz, S., Hofner, P., \& \'{A}lvarez,
C. V. 2004, \apjs, 155, 149
\bibitem[La Vigne et al.(2006)]{La:06} La Vigne, M. A., Vogel, S.
N., \& Ostriker, E. C. 2006, \apj, 650, 818
\bibitem[Liszt(1985)]{Liszt:85} Liszt, H. S. 1985, in IAU Symp. 106,
The Milky Way Galaxy, ed. H. van Woerden et al., p.283
\bibitem[Loinard et al.(2008)]{Loinard:08} Loinard, L., Torres, R.
M., Mioduszewski, A. J., \& Rodr\'{\i}guez, L. F. 2008, \apj, 675L, 29
\bibitem[Loren(1989)]{Loren:89} Loren, R. B. 1989, \apj, 338, 902
\bibitem[Ma et al.(1998)]{Ma:98} Ma, C., Arias, E. F., Eubanks, T.
M., et al. 1998, \aj, 116, 516
\bibitem[Maucherat(1975)]{Maucherat:75} Maucherat, A. J. 1975, \aap,
45, 193
\bibitem[Mao et~al.(2002)]{Mao:02} Mao, R. Q., Yang, J., Henkel, C.,
\& Jiang, Z. B. 2002, \aap, 389, 589
\bibitem[Menten et al.(2007)]{Menten:07} Menten, K. M., Reid, M. J.,
Forbrich, J., \& Brunthaler, A. 2007, \aap, 474, 515
\bibitem[Molinari et al.(1996)]{Mol:96} Molinari, S., Brand, J., Cesaroni,
R., \& Palla, F. 1996, \aap, 308, 573
\bibitem[Molinari et al.(1998)]{Molinari:98} Molinari, S., Brand,
J., Cesaroni, R., Palla, F., \& Palumbo, G. G. C. 1998, \aap, 336, 339
\bibitem[Morgan et al.(1952)]{Mor:52} Morgan, W. W., Sharpless, S.,
\& Osterbrock, D. 1952, \aj, 57, 3
\bibitem[Morgan et al.(1953)]{Mor:53} Morgan, W. W., Whitford, A.
E., \& Code, A. D. 1953, \apj, 118, 318
\bibitem[Moscadelli et al.(2009)]{Mos:09} Moscadelli, L., Reid, M.
J., Menten, K. M., et al. 2009, \apj, 693, 406
\bibitem[Moscadelli et al.(2011)]{Mos:11} Moscadelli, L., Cesaroni,
R., Rioja, M. J., Dodson, R., \& Reid, M. J. 2011, \aap, 526, 66
\bibitem[Muller et al.(2007)]{Muller:07} Muller, S., Dinh-V-Trung,
Lim, J., Hirano, N., Muthu, C., \& Kwok, S. 2007, \apj, 656, 1109
\bibitem[Nagayama et al.(2011)]{Nagayama:11} Nagayama, T., Omodaka,
T., Nakagawa, A., et al. 2011, \pasj, 63, 23
\bibitem[Oort et al.(1958)]{Oort:58} Oort, J. H., Kerr, F. J., \&
Westerhout, G. 1958, \mnras, 118, 379
\bibitem[Palla et al.(1993)]{Palla:93} Palla, F., Cesaroni, R., Brand,
J., et al. 1993, \aap, 280, 599
\bibitem[Patsis et al.(1997)]{Patsis:97} Patsis P. A., Grosbol P.,
\& Hiotelis N. 1997, \aap, 323, 762
\bibitem[Petrov et al.(2005)]{Petrov:05} Petrov, L., Kovalev, Y.~Y.,
Fomalont, E., \& Gordon, D. 2005, \aj, 129, 1163
\bibitem[Plume et al.(1992)]{Plume:92} Plume, R., Jaffe, D. T., \&
Evans, II, N. J. 1992, \apjs, 78, 505
\bibitem[Racine(1968)]{Racine:68} Racine, R. 1968, \aj, 73, 233
\bibitem[Reid et al.(2009a)]{Reid:09a} Reid, M. J. , Menten, K. M.,
Brunthaler, A., et al. 2009a, \apj, 693, 397
\bibitem[Reid et al.(2009b)]{Reid:09b} Reid, M. J., Menten, K. M.,
Zheng, X. W., et al. 2009b, \apj, 700, 137
\bibitem[Russeil(2003)]{Rus:03} Russeil, D. 2003, \aap, 397, 133
\bibitem[Russeil et al.(2007)]{Rus:07} Russeil, D., Adami, C., \&
Georgelin, Y. M. 2007, \aap, 470, 161
\bibitem[Rygl et al.(2010)]{Rygl:10} Rygl, K. L. J., Brunthaler, A.,
Reid, M. J., et al. 2010, \aap, 511, 2
\bibitem[Rygl et al.(2012)]{Rygl:12} Rygl, K. L. J., Brunthaler, A.,
Sanna, A., et al. 2012, \aap, 539, 79
\bibitem[Schneider et al.(2007)]{Schneider:07} Schneider, N., Simon,
R., Bontemps, S., Comer$\acute{o}$n, F., \& Motte, F. 2007, \aap,
474, 873
\bibitem[Schneider et al.(2002)]{Schneider:02} Schneider, N., Simon,
R., Kramer, C., Stutzki, J., \& Bontemps, S. 2002, \aap, 384, 225
\bibitem[Sch\"{o}nrich et al.(2010)]{Schonrich10} Sch\"{o}nrich,
R., Binney, J. J., \& Dehnen, W. 2010, \mnras, 403, 1829
\bibitem[Scoville et al.(2001)]{Scoville:01} Scoville, N. Z., Polletta,
M., Ewald, S., Stolovy, S. R., Thompson, R., \& Rieke, M. 2001, \aj,
122, 3017
\bibitem[Shevchenko \& Yabukov(1989)]{Shevchenko:89} Shevchenko,
V. S., \& Yabukov, S. D. 1989, \sovast, 33, 370
\bibitem[Slysh et al.(1997)]{Slysh:97} Slysh, V. I., Dzura, A. M.,
Val'tts, I. E., \& Gerard, E. 1997, \aaps, 124, 85
\bibitem[Smith et al.(2001)]{Smith:01} Smith, N., Jones, T. J., Gehrz,
R. D., Klebe, D., \& Creech-Eakman, M. J. 2001, \aj, 121, 984
\bibitem[Snell \& Edwards(1981)]{Snell:81} Snell, R. L., \& Edwards,
S. 1981, \apj, 251, 103
\bibitem[Steiman-Cameron(2010)]{Steiman:10} Steiman-Cameron, T. Y.
2010, in a conf. in honour of K. C. Freeman, Galaxies and their Masks,
ed. D. L. Block et al., p. 45
\bibitem[Takakuwa et al.(2007)]{Tak:07} Takakuwa, S., Ohashi, N.,
Bourke, T. L., et al. 2007, \apj, 662, 431
\bibitem[Taylor \& Cordes(1993)]{Taylor:93} Taylor, J. H., \& Cordes
J. M. 1993, \apj, 411, 674
\bibitem[Trinidad et al.(2004)]{Trinidad:04} Trinidad, M. A., Curiel,
S., Torrelles, J. M., et al. 2004, \apj, 613, 416
\bibitem[Urasin(1987)]{Ura:87} Urasin, L. A. 1987, Soviet Astronomy
Letters, 13, 356
\bibitem[Vall\'{e}e(2002)]{Val:02} Vall\'{e}e, J. P. 2002, \apj, 566,
261
\bibitem[van der Walt et al.(1996)]{Walt:96} van der Walt, D. J.,
Retief, S. J. P., Gaylard, M. J., \& MacLeod, G. C. 1996, \mnras,
282, 1085
\bibitem[Weintraub et al.(1994)]{Weintraub:94} Weintraub, D. A.,
Kastner, J. H., \& Mahesh, A. 1994, \apj, 420, l87
\bibitem[Wiseman \& Ho(1998)]{Wiseman:98} Wiseman, J. J., \& Ho, P.
T. P. 1998, \apj, 502, 676
\bibitem[Wouterloot \& Brand(1989)]{Wouterloot:89} Wouterloot, J.
G. A., \& Brand J. 1989, \aaps, 80, 149
\bibitem[Wood \& Churchwell(1989a)]{Wood:89a} Wood, D. O. S., \&
Churchwell, E. 1989a, \apj, 340, 265
\bibitem[Wood \& Churchwell(1989b)]{Wood:89b} Wood, D. O. S., \&
Churchwell, E. 1989b, \apjs, 69, 831
\bibitem[Xu et al.(2006a)]{Xu:06a} Xu, Y., Reid, M. J., Menten, K.
M., \& Zheng, X. W. 2006a, \apjs, 166, 526
\bibitem[Xu et al.(2006b)]{Xu:06b} Xu, Y., Shen, Z.-Q., Yang, J.,
et al. 2006b, \aj, 132, 20
\bibitem[Xu et al.(2009)]{Xu:09} Xu, Y., Reid, M. J., Menten, K. M.,
et al. 2009, \apj, 693, 413
\bibitem[Yang et al.(2002)]{Yang:02} Yang, J., Jiang, Z. B., Wang,
M., Ju, B. G., \& Wang, H. C. 2002, \apjs, 141, 157
\bibitem[Yuan(1969)]{Yuan:69} Yuan, C. 1969, \apj, 158, 871
\bibitem[Zhang et al.(2012a)]{Zhang:12a} Zhang, B., Reid, M. J., Menten,
K. M., \& Zheng, X. W. 2012a, \apj, 744, 23
\bibitem[Zhang et al.(2012b)]{Zhang:12b} Zhang, B., Reid, M. J., Menten,
K. M., Zheng, X. W., \& Brunthaler, A. 2012b, \aap, 544A, 42

\end{thebibliography}
\end{document}